\documentclass[11pt,journal,onecolumn]{IEEEtran}

\usepackage{amsmath,url}
\usepackage{amssymb,amsbsy,verbatim,array}
\usepackage{epsfig,pstricks,psfrag,theorem,cite,graphicx,xcomment}

\let\intern=\iffalse

\def\argmax{\operatorname{arg~max}}
\def\figref#1{Fig.\,\ref{#1}}%
\def\E{\mathbb{E}}

\def\R{\mathbb{R}}
\def\Z{\mathbb{Z}}
\def\N{\mathbb{N}}
\def\W{\mathcal{W}}
\def\K{\mathcal{K}}

\def\calL{\mathcal{L}}

\def\L{\mathbb{L}}
\def\Lo{\L^{!o}}

\def\C{\mathbb{C}}
\def\calC{\mathcal{C}}
\def\calT{\mathcal{T}}
\def\c{c_\mathrm{ex}}
\def\ie{{\em i.e.}}
\def\eg{{\em e.g.}}

\def\var{\operatorname{var}}

\def\sir{\mathrm{SIR}}

\def\T{\mathbb{T}}

\def\d{\mathrm{d}}
\def\opt{\mathrm{opt}}

\def\one{\mathbf{1}}

\def\comment#1{{\bf ($\Rightarrow$ #1 $\Leftarrow$)}}

\newtheorem{theorem}{Theorem}

\newtheorem{lemma}{Lemma}
\newtheorem{corollary}[theorem]{Corollary}



\let\figs=\iftrue

\begin{document}
\title{Interference in Lattice Networks}
\author{Martin Haenggi
\thanks{M. Haenggi is with the University of Notre Dame, IN, USA, {\small\tt mhaenggi@nd.edu}.  This work has been supported
by the NSF (grants CNS 04-47869, CCF 728763) and the DARPA/IPTO IT-MANET program
(grant W911NF-07-1-0028). Manuscript date: \today.}}

\maketitle
\begin{abstract}
Lattices are important as models for the node locations in wireless networks for two
main reasons:
(1) When network designers have control over the placement of the nodes, they
often prefer a regular arrangement in a lattice for coverage and interference reasons.
(2) If nodes are randomly distributed or mobile, good channel access schemes ensure
that concurrent transmitters are regularly spaced, hence the locations of the transmitting nodes are well approximated
by a lattice. In this paper, we introduce general interference bounding techniques that permit the
derivation of tight closed-form upper and lower bounds for all lattice networks, and we present and analyze 
optimum or near-optimum channel access schemes for one-dimensional, square, and triangular
lattices.
\end{abstract}
\begin{IEEEkeywords}
Wireless networks, interference, time-division multiple-access, geometry
\end{IEEEkeywords}
\section{Introduction}
\subsection{Motivation and contributions}
Wireless networks where nodes are arranged regularly in a lattice have advantages
in terms of coverage, for example in sensor networks or for cellular base stations, and
in terms of interference, since it is much easier to devise good channel access schemes
than in networks where nodes are randomly deployed or mobile. 
Despite this advantage, relatively little work has focused on the interference characterization in such
networks. Furthermore, interference results for lattices also provide bounds on optimally
scheduled general wireless networks, since the goal of scheduling is to maximize the
spacing between a receiver and its interfering transmitters, while maintaining a certain
density of transmitters.

In this context, this paper makes three contributions:
\begin{itemize}
\item We introduce general bounding techniques for the interference in lattice networks.
\item We apply these bounds to {\em transmitter-centric} MACs (MAC schemes that schedule
transmitters without considering the location of their respective receivers, such as CSMA-type scheduling
without RTS/CTS). In this case, transmitters form a lattice, but interference has to be measured at the
receiver, where the interference is necessarily larger than if it were measured at the location of the
desired transmitter. The transmitter-receiver distance $r$  results in {\em excess interference}.
We show that quadratic approximations of the form $I(r)\approx I(o)+\c r^2$, where $I(o)$ is the
interference at the desired transmitter and $\c r^2$ is the excess interference due to the offset of the
receiver, are highly accurate for small $r$. These results are relevant for well scheduled wireless networks with
arbitrary node distribution.
\item  For networks where all nodes form a lattice, we analyze and compare the interference of
different TDMA schedulers, and we provide schemes that are very close to optimum (if not
optimum) for one-dimensional, square and triangular networks. In the one-dimensional case,
we also provide results on the achievable rate and the transport capacity.
\end{itemize}

\subsection{Related work}
While a growing body of work studies interference in random networks (see, \eg, \cite{net:Haenggi08now,net:Haenggi09jsac} and
references therein), only few papers have addressed the issue of interference in lattice networks.
In \cite{net:Hekmat04wireless}, bounds on the interference in triangular networks were derived using a
relatively crude upper bound on the Riemann zeta function that is within 25\% of the true value for the
range of $2$ to $4$. We will derive a much tighter bound that is within 1.3\%.
A TDMA scheduling scheme for square lattices that is optimum for the case
where the density of concurrent transmitters is $1/4$ is suggested in \cite{net:Liu05globecom}.
Here, we provide near-optimum scheduling schemes for
any density $1/m^2$, $m\in \N$.
The interference distribution in one-dimensional networks with Rayleigh fading is analyzed in
\cite{net:Mathar95wn} for the case where all nodes transmit, and 
\cite{net:Haenggi09twc} derives outage results an throughput-optimum TDMA schedulers for the 
same type of network. Finally, the single-hop throughput for two-dimensional lattice networks with Rayleigh
fading is approximated in \cite{net:Liu05eurasip}. For non-fading channels, \cite{net:Hong07eurasip} provides
throughput results for general TDMA schemes in two-dimensional lattice networks. The interference is expressed
using complicated infinite double sums (that are evaluated numerically), for which we will present tight bounds.

The network models in this paper are entirely deterministic, although, of course, the interference results derived 
correspond to the expected interference in fading channels.

The paper is organized as follows. We first introduce the bounding techniques (Section II), followed by
three application sections that discuss one-dimensional networks (Section III),
square lattice networks (Section IV), and triangular networks (Section V). Conclusions are drawn in
Section IV.

\section{Bounding Techniques}
\label{sec:techniques}
In this section we introduce the basic techniques that will be used to bound the interference.
\subsection{Upper bounds}
Let $\L\subset\R^d$ be a $d$-dimensional lattice, \ie,
\begin{equation}
   \L\triangleq\{x=\mathbf{G}u\colon u\in\Z^d\}\,, 
   \label{lattice}
\end{equation}
where $\mathbf{G}\in \R^{d\times d}$ is the generator matrix. It is assumed that $\det\mathbf{G}\neq 0$ to exclude degenerate cases.
Important cases include the square integer and the
triangular lattice in two dimensions, both with nearest-neighbor distance $1$:
\begin{equation}
   \mathbf{G}_\mathrm{sq}=
\begin{bmatrix} 1& 0\\0 & 1 \end{bmatrix}\,;\qquad \mathbf{G}_\mathrm{tri}=
\begin{bmatrix} 1 &1/2\\0 & \sqrt{3}/2 \end{bmatrix} 
\label{generators}
\end{equation}
The lattice $\L$ has the properties that it includes the origin $o=(0,\ldots,0)$ and that each lattice point is
centered in its Voronoi cell, \ie, if $V_\L(x)$ is the Voronoi cell of lattice point $x\in\L$ and $U_x$ is a uniformly randomly distributed
random variable on $V_\L(x)$, then $\E U_x =x$. The volume of each Voronoi cell is $V=|\det\mathbf G|$, and the
density of the lattice is $\lambda=V^{-1}$ (points per unit volume).

Let $\Lo\triangleq\L\setminus\{o\}$\footnote{The superscript $^{!o}$ is borrowed from stochastic geometry, where it denotes a
random point set, conditioned on having a point at the origin but excluding that point.}. Then the interference at point $z$ is defined as
\[ I(z)\triangleq\sum_{x\in\Lo} \ell(x-z)\,, \]
where $\ell(x)\colon\R^d\rightarrow \R^+$ is the path loss function, assumed monotonically decreasing and isotropic, \ie, there exists a
function $\ell'(r)\colon \R^+\rightarrow \R^+$ such that $\ell'(\|x\|)\equiv \ell(x)$, where $\|\cdot\|$ is the standard Euclidean norm.
 Further assume
that $\ell'(r)=o(r^{-d})$ as $r\rightarrow\infty$ (otherwise the interference is infinite)
and that $\ell'(r)$ has a {\em convex tail}, \ie, there is a finite radius, defined as
\[ r_c\triangleq\inf \{ r>0 \mid \ell'(\rho)\text{ convex for } \rho\geqslant r\} \,.\]
All commonly and rarely used path loss functions satisfy these properties.
The origin is assumed to be the desired transmitter, so it does not contribute to the interference. The distance
$\|z\|$ is restricted to values that ensure that $z$ is not too close to an interferer, \ie, $\|z\|<\|g_i\|$, for $1\leq i\leq d$, where
$g_i$ are the column vectors that constitute $\mathbf{G}$.

The first theorem exploits the tail convexity of the path loss function to yield an upper bound on the interference.
Let $b_x(r)$ be the $d$-dimensional ball of radius $r$ centered at $x$.
\begin{theorem}[Voronoi upper bound]
\label{thm:voronoi}
Without offset: Let $L$ be a subset of $\Lo$ such that there is no overlap between any of the Voronoi cells
of the points in $L$ and $b_o(r_c)$. Then
\begin{equation}
   I(o)\leqslant \sum_{x\in\Lo\setminus L} \ell(x)+\frac1V\sum_{x\in L} \int_{V_\L(x)} \ell(y)\d y \,.
 \label{voronoi}
\end{equation}
With offset: Let  $L$ be a subset of $\Lo$ such that
$L$ does not include any points whose Voronoi cells overlap with $b_o(r_c+\|z\|)$. Then
\begin{equation}
   I(z)\leqslant \sum_{x\in\Lo\setminus L} \ell(x-z)+\frac1V\sum_{x\in L} \int_{V_\L(x)} \ell(y-z)\d y \,.
 \label{voronoi2}
\end{equation}
\end{theorem}
\begin{IEEEproof}
We only need to prove \eqref{voronoi2}. Since $\ell'(r)$ is convex for $r>r_c$ and the Voronoi cells of all points in $L$ lie outside $b_o(r_c+\|z\|)$, it follows from
Jensen's inequality that $\ell(x-z)=\ell(\E(U_x)-z)\leqslant \E(\ell(U_x-z))$ for all $x\in L$, where $U_x$ is uniformly distributed 
on $V_\L(x)$.
\end{IEEEproof}
Next we state a corollary that applies to two-dimensional lattices and provides a bound that is simple to evaluate, as it is based
on a radial outer bound on the integration region.
\begin{corollary}[Radial bound for two-dimensional lattices]
\label{cor:radial}
Let $\rho_x=\min_{y\in V_\L(x)} \|y\|$ be the (minimum) distance of the Voronoi cell of point $x$ to the origin, and
define $\mathcal{R}_z\triangleq \{\rho_x\colon x\in\Lo,\;\rho_x\geqslant r_c+\|z\| \}$.
Without offset: For any $r_b\in\mathcal{R}_o$,
\begin{equation}
      I(o) < \sum_{\substack{x\in\Lo\\\rho_x< r_b}}\ell(x)+\frac{2\pi}{V} \int_{r_b}^\infty r\ell'(r)\d r \,. 
 \label{radial}
\end{equation}
With offset: For any $r_b\in\mathcal{R}_z$,
\begin{equation}
I(z)<\sum_{\substack{x\in\Lo\\\rho_x< r_b}}\ell(x-z)+\frac{2\pi}{V} \int_{r_b-\|z\|}^\infty r\ell'(r)\d r \,.
\label{radial_offset}
\end{equation}
\end{corollary}
\begin{IEEEproof}
No offset: Since the union of all Voronoi cells of the points not included in the sum is a strict subset of $\R^2\setminus b_o(r_b)$, 
\eqref{radial} follows from Theorem \ref{thm:voronoi}. With offset: The change from $o$ to $z$ means that $\ell(x)$ is to be replaced
by $\ell(x-z)$ in the sum and the integral. The integral in \eqref{radial} is taken over $\R^2\setminus b_o(r_b)$, thus, with the offset,
the domain of integration is $\R^2\setminus b_{-z}(r_b)$. Since this integral may be tricky to calculate, we replace it conservatively
with $\R^2\setminus b_o(r_b-\|z\|)$ for the upper bound.
\end{IEEEproof}
{\em Remarks:}
\begin{enumerate}
\item For the case where the path loss law $\ell(x)$ has a power law tail with exponent $\alpha$,
the integral evaluates to $2\pi b^{2-\alpha}/(\alpha-2)$, where $b$ is the lower integration bound.
\item The reason why $r_b$ is restricted to the set $\mathcal{R}$ is that if $r_b$ is not the smallest distance of a Voronoi
cell, then $r_b$ can always be increased to the next larger smallest distance without changing the set over which the
sum is taken, thereby yielding a better bound. In other words, for any set $L$ chosen in Theorem \ref{thm:voronoi}, the
corresponding $r_b$ here should be the largest radius such that $b_o(r_b)$ does not overlap with any of the Voronoi
cells of the points in $L$.
\item The integral in \eqref{radial} can be interpreted as the mean interference stemming from a Poisson point process
of intensity $\lambda(r)=\one_{\{r>r_o\}}/V$. As one would expect, outside of a certain radius from the origin, only the
intensity of the point process matters, rather than its higher-order statistics (such as second-moment measures). 
The corollary provides a bound on this radius such that  the interference in the lattice is upper bounded by the interference
in the Poisson point process of the same intensity.
\end{enumerate}

\subsection{Lower bounds}
To obtain lower bounds, the distances of two or more points (in the convex region of the path loss function) 
are replaced by their averages.  Evaluating the path loss at this average point, multiplied by the number of points that
the average was taken over, yields a {\em lower} bound. This technique is not restricted to lattices.
\begin{theorem}[Lower bound for general point sets]
\label{thm:lower}
Let $\L\subset\R^d$ be an arbitrary discrete set of points that includes the origin, and let
$\Lo\triangleq\L\setminus\{o\}$.
For any $L\subset \L\cap b_o^c(r_c+\|z\|)$, let $\bar x_L=(1/|L|) \sum_{x\in L}x$ be the average of the
points in $L$. Then
\begin{equation}
I(z) \geqslant \sum_{x\in\Lo\setminus L} \ell(x-z)+|L| \ell(\bar x_L-z) \,.
\end{equation}
\end{theorem}
\begin{IEEEproof}
This follows again from Jensen's inequality. Moving any set of lattice points (outside distance $r_c$ from the origin)
to their average yields a lower bound due to the convexity of $\ell$ in this regime.
\end{IEEEproof}
In general, this theorem will be applied repeatedly; for example, many pairs of points are formed and replaced by their averages. Better
bounds can be expected if $|L|$ is kept small, and if the points in $L$ are located nearby.

\subsection{Bounds on zeta function}
For power law path-loss functions, the interference in lattice networks can often be expressed using the standard zeta
function (without offset) or the generalized zeta function (with offset).
The following lemma provides tight yet simple
closed-form bounds on the standard and generalized Riemann zeta function.
\begin{lemma}
\label{lem:zeta}
The generalized Riemann zeta function\footnote{also called Hurwitz zeta function}
\[ \zeta(\alpha,1-z)\triangleq \sum_{k=1}^\infty (k-z)^{-\alpha} \,,\qquad \alpha>1,\, |z|<1\,, \]
is tightly upper bounded by
\begin{equation}
   \zeta(\alpha,1-z) \lessapprox (1-z)^{-\alpha}+\frac{\left(\frac 32-z\right)^{1-\alpha}}{\alpha-1} \,. 
   \label{zeta_bound}
\end{equation}
and lower bounded by
\begin{equation}
  \zeta(\alpha,1-z) > (1-z)^{-\alpha}+\frac{\left(2-z\right)^{1-\alpha}}{\alpha-1} \,. 
   \label{zeta_lower_bound}
\end{equation}
For the standard zeta function $\zeta(\alpha)\equiv \zeta(\alpha,1)$, an alternative upper bound that is
even tighter for $\alpha>2$ is
\begin{equation}
  \zeta(\alpha)\lessapprox \frac{\alpha-1+2^{-\alpha}}{\alpha-1-(\alpha-1)2^{-\alpha}}\,,
  \label{zeta_bound2}
\end{equation}
and a good lower bound is
\begin{equation}
 \zeta(\alpha)\gtrapprox \frac{6^\alpha}{6^\alpha-3^\alpha-2^\alpha-1} \,.
 \label{zeta_lower}
\end{equation}
\end{lemma}
\begin{IEEEproof}
The first bound \eqref{zeta_bound} is an application of Theorem \ref{thm:voronoi} with $\L=\N$ and $L=\N\setminus\{1\}$.
Let $X_k$ be a random variable that is uniformly randomly distributed in $[k-1/2-z,k+1/2-z)$.
From Jensen's inequality, it follows that $(\E X_k)^{-\alpha}< \E(X_k^{-\alpha})$ for all $k>1/2+z$ and $\alpha>0$. 
Expressing the zeta function as
\[ \zeta(\alpha,1-z)=
  \sum_{k=1}^n (k-z)^{-\alpha}+\sum_{k=n+1}^\infty (\E X_k)^{-\alpha} \,\]
  for arbitrary $n\in\N$
and upper bounding each term $(\E X_k)^{-\alpha}$ by $ \E(X_k^{-\alpha})$,  we obtain the bounds
\[ \zeta(\alpha,1-z) < \sum_{k=1}^n (k-z)^{-\alpha} + \int_{n+1/2-z}^\infty x^{-\alpha}\d x\,, \qquad n\in\N \,,\]
  which are increasingly tight as $n$ grows. \eqref{zeta_bound} is the bound for $n=1$. For the lower bound
  \eqref{zeta_lower_bound}, we change the support of all $X_k$ to $[k-z, k+1-z]$. In this case, since $X_k\geqslant k-z$,
  we have $(k-z)^{-\alpha}>\E(X_k^{-\alpha})$
  and thus
  \[ \zeta(\alpha,1-z) > \sum_{k=1}^n (k-z)^{-\alpha} + \int_{n+1-z}^\infty x^{-\alpha}\d x\,, \qquad n\in\N \,.\]
The bound \eqref{zeta_lower_bound} is the bound for $n=1$.
Regarding \eqref{zeta_bound}, we consider the special case  $z=1/2$. It is 
 straightforward to show that
 \begin{equation}
   \zeta(\alpha,1/2)=(2^\alpha-1) \zeta(\alpha) \,,
\end{equation}
which, together with \eqref{zeta_bound} for $z=1/2$ yields $(2^\alpha-1)\zeta(\alpha)\lessapprox 2^{\alpha}+1/(\alpha-1)$, which proves
\eqref{zeta_bound2}. The lower bound \eqref{zeta_lower} stems from an application of Theorem \ref{thm:lower} (with $z=0$) and
\begin{equation}
   \L=\bigcup_{k\in\N} \{-6k-1,-6k+3,6k-3,6k-1\} 
   \label{smart_tdma_set}
\end{equation} 
and averaging on the distances of the points $-6k-1$ and $6k-1$, \ie, replacing them by two points at distance $6k$.
Hence we compare the two sums
\[ S_1=\sum_{k\in\N} (3k-\one_{\{k\text{ even}\}})^{-\alpha}+(3k+\one_{\{k\text{ even}\}})^{-\alpha}\,;\qquad
 S_2=2\sum_{k\in\N} (3k)^{-\alpha} \,,\]
 where $S_2$ is obtained from $S_1$ by averaging over the two arguments $3k-1$ and $3k+1$ whenever $k$ is even. The terms for odd $k$
 are the same in $S_1$ and $S_2$.
 From Theorem \ref{thm:lower}, we have $S_1>S_2$, where
 \[ S_1= \zeta(\alpha)-1+\sum_{k\in\N} (3k)^{-\alpha}-\sum_{k\in\N} (2k)^{-\alpha} -\sum_{k\in\N} (6k)^{-\alpha}
  = \zeta(\alpha)(1+3^{-\alpha}-2^{-\alpha}-6^{-\alpha})-1 \,,\]
  and $S_2=2\cdot 3^{-\alpha}\zeta(\alpha)$. 
\end{IEEEproof}
{\em Remarks:}
\begin{enumerate}
\item For $\alpha<2$ (and $z=0$), the first bound \eqref{zeta_bound} is tighter than the second one \eqref{zeta_bound2}. They are equal at $\alpha=2$, which is the
value of $\alpha$ where the first bound is loosest. The difference to the actual value $\zeta(2)=\pi^2/6$ is $(10-\pi^2)/6\approx 0.0217$.
The two bounds \eqref{zeta_bound2} and \eqref{zeta_lower} are so tight that when plotting both curves over a range $[\alpha_l,\alpha_u]$
with $\alpha_u-\alpha_l>1$ they appear as one. For $\alpha\downarrow 1$, the upper bound is tighter, while for practical
values of $\alpha$, the lower bound is.
\item A simpler and looser lower bound is
\[ \zeta(\alpha) > \frac{2^\alpha-1}{2^\alpha-2} \,,\] 
obtained from summing over $1,2,4,4,6,6,8,8,\ldots$ instead of $\N$.
\end{enumerate}

\section{One-dimensional Lattice Networks}
\subsection{Interference with offset}
Here we assume that the desired transmitter is located at the origin $o$, the receiver at position $0<z<1$,
and interferers at positions $\Z\setminus\{o\}$. 
Unless otherwise noted, we use the standard power path loss law $\ell(x)=\|x\|^{-\alpha}$.
Since $I(z)=\zeta(\alpha,1-z)+\zeta(\alpha,1+z)$, we can apply Lemma \ref{lem:zeta} to obtain a tight closed-form upper bound:
\begin{equation}
 I(z)\lessapprox (1-z)^{-\alpha}+(1+z)^{-\alpha}+\frac{1}{\alpha-1}\left[\left(\frac32-z\right)^{1-\alpha}+\left(\frac32+z\right)^{1-\alpha}\right]
 \label{onedim_approx}
\end{equation}
Due to the symmetry of the arrangement, $I(z)$ is an even function. For integer $\alpha$, the bound is a rational function in $z^2$, with the numerator polynomial of degree $3\alpha-1-\one_{\{\alpha\text{ even}\}}$, and the denominator polynomial of degree $4\alpha-2$. A quadratic Taylor expansion at $z=0$ gives a good approximation for small $z$ (in this case, $z<1/4$ is small enough for a very good approximation):
\[ I(z)\approx 2\zeta(\alpha)+\c z^2\,,\]
where $\c\triangleq I''(0)=(1/2) \d^2 I(z)/(\d z)^2|_{z=0}$ is the (transmitter-receiver) {\em offset coefficient}.
We obtain
\begin{equation}
 \c=\alpha^2+\alpha\left(1+\left(\frac23\right)^{\alpha+1}\right) \,. 
 \label{cex}
\end{equation}
For all practical values of $\alpha$, a good lower bound is
$\c > \alpha^2+\alpha$, and a good approximation is $\c\approx \alpha^2+\alpha+1/2$.
It is also possible to give a simple yet accurate approximation for the fourth-order coefficient:
\[ \c^{(4)} \approx \frac{1}{12}\alpha^4+\frac12 \alpha^3+\alpha^2+\frac12 \alpha \]
This approximation for $\c^{(4)}$ is within $1\%$ for all $\alpha\geqslant 1$.
For practical $\alpha$ and small $z$, the quadratic term always dominates the fourth-order one.

Eqn.~\eqref{onedim_approx} can be used to find the displacement $z$ that maximizes the {\em transport capacity}
$T(z)=z\log_2(1+z^{-\alpha}/I(z))$, where $z$ is the link distance, and the logarithmic term is the rate $R(z)$.
Since $1+z^{-\alpha}/I(z)$ is a rational function for integer $\alpha$, a series
expansion of the form $R(z)\approx c_1+c_2\log z+c_3 z^2$ can easily be found. Multiplying by $z$ and setting the
derivative to zero leads to solving an equation of the form $1-2 \log z/a +b z^2=0$, which yields
$z_\opt=\exp(a/2-\mathcal{W}(ab e^a))/2)$, where $\mathcal{W}$ is the Lambert W function. For $\alpha=2$, this 
yields $z_\opt=0.223$, while a numerical investigation yields $z_\opt=0.224$. For $\alpha=4$, the gap is slightly
larger: The analytical result is $z_\opt=0.217$, and the numerical one is $z_\opt=0.222$. For non-integer $\alpha$,
similar series expansions can be derived by substituting $x=z^\alpha$ and optimizing $x$. It can be concluded that
the optimum value of $z$ is between $0.2$ and $0.25$ and rather insensitive to the value of $\alpha$.

\subsection{Application to TDMA scheduling patterns}
\subsubsection{Separating transmitters}
Let the nodes at positions $m\Z$ be transmitters, each one transmitting to the next node on the right.
Focusing on the desired transmitter at the origin and its receiver at location $1$, the other transmitters
at positions $\K=m\Z\setminus\{o\}$ are interferers. 
This is an $m$-phase TDMA pattern, since it will require $m$ time slots
to give each node a transmission opportunity. The interference at location $1$ can be expressed as
\[ I_m=\sum_{k\in\K} |k-1|^{-\alpha}=m^{-\alpha} I(1/m)\,,\qquad m>1\,. \]
With Lemma \ref{lem:zeta} and
\[ G(m,b)\triangleq \frac{(m-1)^\alpha+(m+1)^\alpha}{(m^2-1)^\alpha}+
    \frac{(bm+1)^{\alpha-1}+(bm-1)^{\alpha-1}}{m(\alpha-1)(b^2m^2-1)^{\alpha-1}}\,,\]
    we have
 \[ G(m,2)<I_m<G(m,3/2) \,. \]
The maximum achievable rate is 
\begin{equation}
  R_m=\log_2\left(1+I_m^{-1}\right)\approx \log_2\left(1+\frac{m^{\alpha+2}}{2\zeta(\alpha)m^2+\c}\right) \,,
\end{equation}
where $\c$ is given by \eqref{cex}, and the throughput is $T_m=R_m/m$.
The derivative $\partial T_m/\partial m$ indicates that the optimum $m$ is always $m=4$ or $m=5$, with
$T_4\approx T_5$ for all $\alpha\in(1,20]$. 
Due to the smaller power consumption, $m=5$ is preferred in practice even if $m=4$ yielded a slightly higher throughput.
So, interestingly, irrespective of the path loss exponent, $m=5$ is an excellent choice --- assuming the rate of
transmission is adjusted to the interference present for that path loss exponent. A first-order Taylor expansion at $\alpha=2$
yields $T_5\approx 0.60+0.53(\alpha-2)$, which is a very good approximation for the practical range of $\alpha$.

The throughput $R_m/m$ is identical to the transport capacity $z R(z)$ considered in the previous subsection when $z=1/m$.
Thus the result that $m=4$ or $m=5$ are optimum for all $\alpha$ also follows from the fact that
$z_\opt$ is between $1/4$ and $1/5$ for all $\alpha$. 

The optimum $m$ for one-dimensional networks with Rayleigh fading was derived in \cite{net:Haenggi09twc}. It was found
that $m_\opt=3$ for $\alpha>2$. So in the fading case, transmitters can be packed more closely.

This $m$-phase TDMA scheme suffers from an imbalance in the interference from the left and right; nonetheless, it is optimum for unidirectional traffic. For bidirectional traffic, a better TDMA scheme
exists, as discussed next.

\subsubsection{Balanced TDMA schemes for bidirectional traffic}

Here we analyze TDMA schemes for bidirectional traffic, which achieve higher throughput than in the unidirectional case,
since it is possible to balance interference from the two nearest interferers.
These MAC schemes separate the receivers from their interfering transmitter in an optimum way.
In other words, instead of placing transmitters optimally, these schemes place transmit-receiver pairs
optimally.

We use the following notation to define transmission patterns: Transmitters are indicated by T, with an
arrow on top to indicate on which side their receiver sits, and receivers are denoted by R. As a shortcut,
R$^k$ indicates a sequence of $k$ receiving nodes. The period
of the transmission scheme is indicated by a subscript. For example, the unidirectional TDMA scheme with
$m=3$ is described as $(\overrightarrow{\text{T}}$R$\,$R$)_3$ or $(\overrightarrow{\text{T}}$R$^2)_3$, and
for general $m$,  $(\overrightarrow{\text{T}}$R$^{m-1})_m$.

In the balanced schemes, transmitters with their receivers to the right alternate with transmitters with their
receivers to the left. For $m=3$, for example, the pattern is $(\overrightarrow{\text{T}}$R$\,$R$\,$R$\overleftarrow{\text{T}}$R$)_6$.
This has a period of 6, but since two nodes transmit in a group  of 6 nodes, it take 3 time slots for each node to transmit once.
For general $m$, the scheme is $(\overrightarrow{\text{T}}$R$^{m}\overleftarrow{\text{T}}$R$^{m-2})_{2m}$.
These balanced transmission patterns are illustrated in \figref{fig:smart_tdma}.

\figs
\begin{figure}
\centerline{\epsfig{file=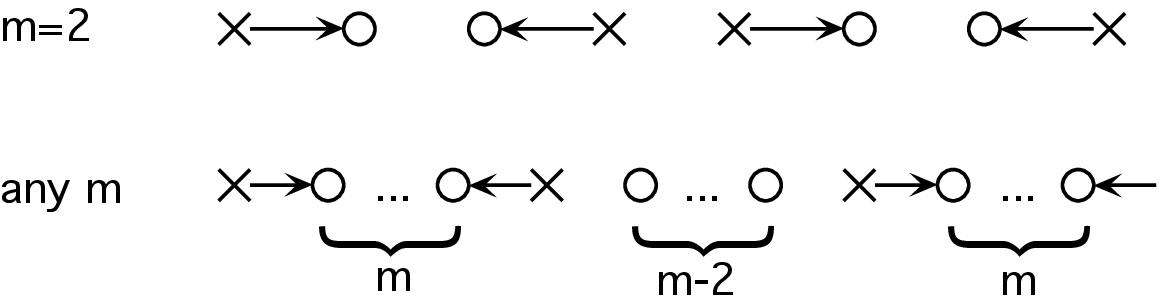,width=9cm}}
\caption{Balanced TDMA schemes for $m=2$ and general $m$. The two nearest interferers are at distances $m$ (to the left) and $m$ (to the right).}
\label{fig:smart_tdma}
\end{figure}
\fi

Let $m=2$ and the receiver under consideration be located at the origin and listening to the transmitter at $-1$, \ie, we focus on the
underlined receiver in the pattern $(\overrightarrow{\text{T}}$\underline{R}$\,$R$\overleftarrow{\text{T}})_4$.
The interferers are located at 
\[ \K_2=\bigcup_{k\in\N} \{-4k-1,-4k+2,4k-1,4k-2\}\,, \]
and the interference is
\[ I_2=\sum_{k\in\K} |k|^{-\alpha}=\zeta(\alpha)-1+\sum_{k\in\N} (2k)^{-\alpha}-2\sum_{k\in\N} (4k)^{-\alpha} 
  = \zeta(\alpha)(1+2^{-\alpha}-2\cdot 4^{-\alpha})-1\,. \]
  $I_2$ can be bounded by bounding the zeta function using Lemma \ref{lem:zeta}. Alternatively, a lower bound is obtained by replacing the
  interferers at positions $4k-1$ and $-4k-1$ by two at position $4k$ and applying
  Theorem \ref{thm:lower}:  
\[ I_2=\sum_{k\in\K} |k|^{-\alpha} > 2\sum_{k\in\N} (2k)^{-\alpha} = 2^{1-\alpha} \zeta(\alpha) >
\frac{2\cdot 3^{\alpha}}{6^{\alpha}-3^{\alpha}-2^{\alpha}-1} \]
For $m=3$,
\[ \K_3=\bigcup_{k\in\N} \{-6k-1,-6k+3,6k-3,6k-1\}\,. \]
The exact expression is
\[ I_3=\sum_{k\in\K} |k|^{-\alpha} = \zeta(\alpha)-1+\sum_{k\in\N} (3k)^{-\alpha}-(2k)^{-\alpha}+ (6k)^{-\alpha} 
  = \zeta(\alpha)(1+3^{-\alpha}-2^{-\alpha}-6^{-\alpha})-1 \,.\]
  Using the lower bound for $\zeta$ from Lemma \ref{lem:zeta}, we obtain
  \[ I_3 > \frac{2^{\alpha+1}}{6^\alpha-3^\alpha-2^\alpha-1} \,.\]
Generalizing to arbitrary $m>1$, there are interferers at $\K_m=\{k\in\N\mid -2km-1,-2km+m,2km-m,2km-1\}$. Replacing
each pair $-2km-1$ and $2km-1$ by two interferers at $2km$, we have
\[ I_m > 2\sum_{k\in\N}(mk)^{-\alpha}=2 m^{-\alpha}\zeta(\alpha) \,,\]
which can be (further) lower bounded by any of the lower bounds on the zeta function from Lemma \ref{lem:zeta}, \eg,
\[ I_m> \frac{2\cdot 6^\alpha}{m^\alpha(6^\alpha-3^\alpha-2^\alpha-1)}
 > \frac{2}{m^\alpha}\left[1+\frac{1}{2^\alpha-2}\right] \,.\]
To find a strict upper bound on $I_m$, we replace the elements $2km-1$ in $\K_m$ by $2km-m+1$, which yields
\begin{align*}
   I_m &\stackrel{(a)}{\leqslant} \sum_{k\in\N} 2(2km-m)^{-\alpha}+(km+1)^{-\alpha}\\
    &=\sum_{k\in\N} 2(2m)^{-\alpha} (k-1/2)^{-\alpha}+m^{-\alpha}(k+1/m)^{-\alpha}\\
    &=2(2m)^{-\alpha}\zeta(\alpha,1/2)+m^{-\alpha}\zeta(\alpha,-1/m)\\
    &\stackrel{(b)}{<} \frac1{m^{\alpha}}\left(2+\frac{2^{1-\alpha}}{\alpha-1}\right)+
       (m+1)^{-\alpha}+\frac1{m(\alpha-1)}\left(\frac{2}{3m+2}\right)^{\alpha-1}\,,
 \end{align*}
 where $(a)$ is strict for $m>2$ and $(b)$ follows from \eqref{zeta_bound} in Lemma \ref{lem:zeta}.
Since the lower bound is tighter and reveals that $I_m$ is essentially proportional to $m^{-\alpha}$, we proceed
with the lower bound to find the throughput-optimum $m$. The SIR is $I_m^{-1}=m^{\alpha}C(\alpha)$, where,
from \eqref{zeta_lower},
\begin{equation}
  C(\alpha)=\frac1{2\zeta(\alpha)}\approx \frac12\left(1-2^{-\alpha}-3^{-\alpha}-6^{-\alpha}\right) \,.
  \label{calpha}
\end{equation}
With $R_m=\log_2(1+m^\alpha C(\alpha))$, the throughput $T_m=R_m/m$ is maximized at
\[ m_\opt^\alpha=-\frac{\W(a)+\alpha}{C(\alpha)\W(a)} \,,\]
(relaxing the integer constraint on $m$), where $a=-\alpha e^{-\alpha}$ and $\W$ is the Lambert W function.
The (real) $m_\opt\in[3,4]$ for $\alpha<7$. Comparing $T_3$ and $T_4$ shows very little difference, and again
power consumption favors $T_4$. A second-order expansion of $T_4$ as a function of $\alpha$ is
$T_4(\alpha)\approx 0.64+0.59(\alpha-2)$. This is better than in the unidirectional case, but the difference
is not large. Noting that $4^\alpha C(\alpha)\approx (4^\alpha-2^\alpha)/2$ for practical $\alpha$, the optimum
rate at $m=4$ is $R_4\approx \alpha-1+\log_2(2^{\alpha}-1)$, or just a bit less than $2\alpha-1$. 
\figref{fig:onedim_through} shows the throughput as a function of $m$ and $\alpha$, together with the approximation
$\log_2(1+m^\alpha C(\alpha))/m$, where $C(\alpha)$ is given in \eqref{calpha}.

\figs
\begin{figure}
\centerline{\epsfig{file=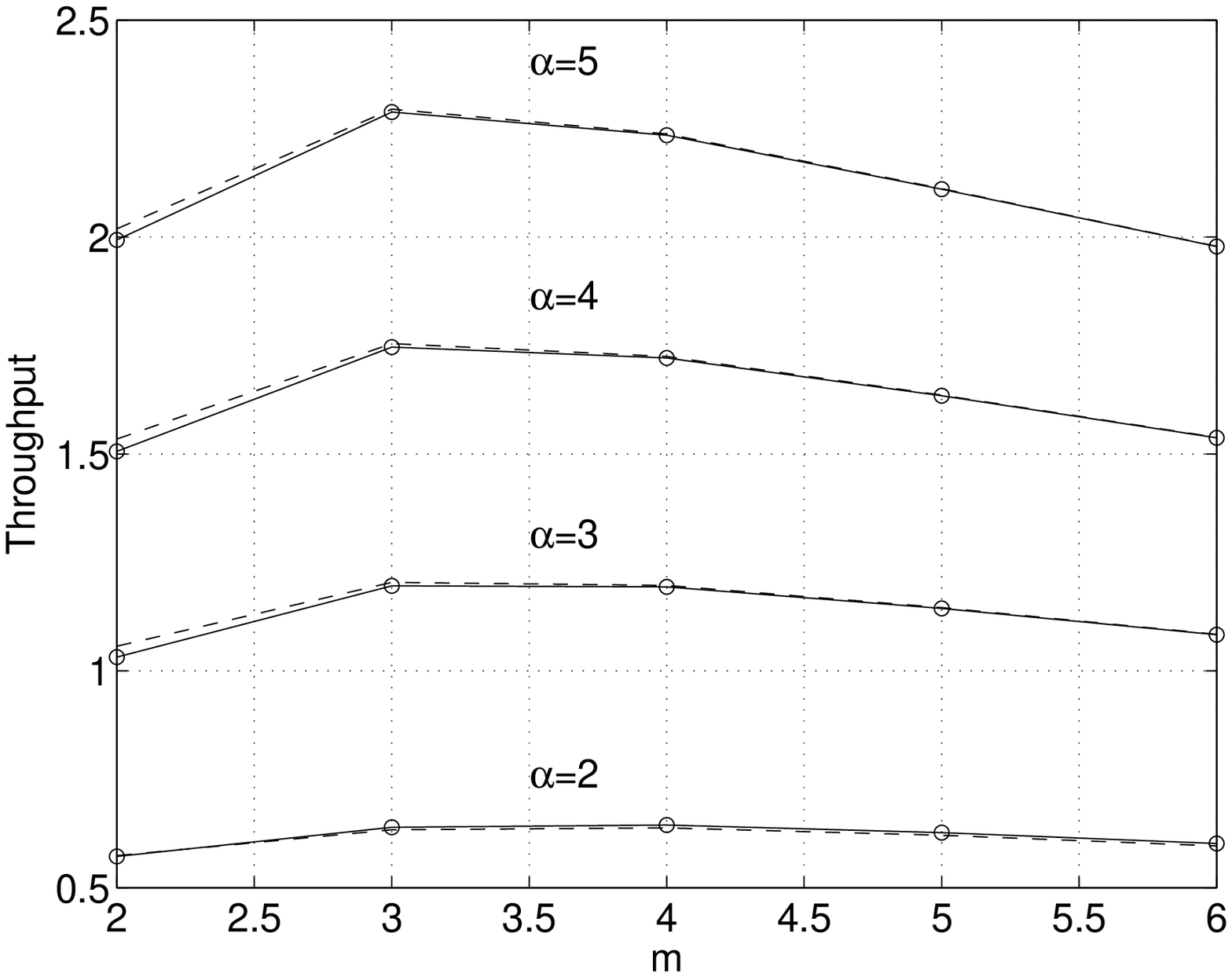,width=9cm}}
\caption{Throughput as a function of path loss exponent $\alpha$ and TDMA parameter $m$ for balanced schemes in
one-dimensional networks. The dashed lines are the approximations $\log_2(1+m^\alpha C(\alpha))/m$.}
\label{fig:onedim_through}
\end{figure}
\fi

\begin{theorem}[Interference density for $\alpha=4$]
The probability density for the interference as $n\rightarrow\infty$ is
\begin{equation}
 f_I(x)=\begin{cases} \displaystyle 4\pi\sum_{i=1}^\infty \frac{(-1)^{i+1} i}{\sinh(i\pi)} i^4\exp(-i^4 x) & \text{ if } x>0 \\
    0 & \text{ if } x=0 \,.\end{cases}
   \label{mathar_limit4}
\end{equation}
\end{theorem}
\begin{IEEEproof}
Using Euler's product formula
\begin{equation}
 \sin(\pi z)\equiv \pi z\prod_{i=1}^\infty \left(1-z^2/i^2\right)\,,\quad z\in \C \,,
 \label{euler_prod}
 \end{equation}
we may write
\begin{align*}
\frac1{a_{\infty,i}}&=\lim_{x\rightarrow i}\prod_{\substack{k=1\\k\neq i}}^\infty (1-x^4/k^4)\\
 &=\lim_{x\rightarrow i} \Bigg[\prod_{\substack{k=1\\k\neq i}}^n (1-i^2/k^2) \prod_{\substack{k=1\\k\neq i}}^n (1+i^2/k^2)\Bigg]\\
 &=\lim_{x\rightarrow i} \frac{\sin(\pi x)\sinh(\pi x)}{(\pi x)^2 (1-(x/i)^4)} \\
 &=\lim_{x\rightarrow i}  \frac{\pi\cos(\pi x)\sinh(\pi x)-\pi\sin(\pi x)\cosh(\pi x)}{2\pi^2(1-(x/i)^4)-\pi^2 x(4x^4/i^4)}\\
 &=\frac{(-1)^{i+1}\sinh(i\pi)}{4\pi i}\,.
\end{align*}
\end{IEEEproof}
Due to the $\sinh$ term, the coefficients $a_{\infty,i}$ decay very quickly, and it is sufficient to consider only the nearest three or even
two interferers. Considering only the nearest interferer yields approximately the right tail of the density, but the probabilities of seeing
little interference are drastically different. This is essentially a diversity effect.

\figref{fig:mathar4} shows the densities $f_{I_n}(x)$ for $n=1,2,3,\infty$. The curve for $n=3$ is virtually indistinguishable
from the limiting case. The mean interference in the infinite case is
$\E(I)=\sum_{i=1}^\infty i^{-4}=\zeta(4)=\pi^4/90$.

\fi

\section{Square Lattices}
In this section, we consider square lattices $\L=\Z^2$.
In the case of the power law path loss, we may use results on lattice sums, see, \eg, \cite{net:Zucker74}, to express the
interference at the origin:
\begin{equation}
  I(o)=\sum_{x\in\Lo} \|x\|^{-\alpha}=4\zeta(\alpha/2)\beta(\alpha/2)\,, 
\label{square_lattice_origin}
\end{equation}
where 
\[ \beta(x)\triangleq \sum_{i=1}^\infty \frac{(-1)^{i+1}}{(2i-1)^x}\, \]
is the Dirichlet beta function. $\beta(2)$ is Catalan's constant $K=0.916$. So, for $\alpha=4$, $I=2\pi^2 K/3\approx 6.03$.
These expressions are not closed-form and are restricted to the origin,
and thus do not provide much insight into the behavior of the interference.
We apply the techniques described in Section \ref{sec:techniques} to derive tight closed-form upper and lower bounds.
The path loss law assumed is again $\ell(x)=\|x\|^{-\alpha}$.

\subsection{Lower interference bounds}
To obtain a good approximation on \eqref{square_lattice_origin}, we group the nodes into (square) ``rings" of
increasing distances. In the $k$-th ring, there are $8k$ nodes at distances between $k$ and $\sqrt 2k$.
Let $\bar D_k$ be the average of the distances of the $k$-th ring of nodes. Then, taking the sum over the nearest
ring separately,
\[ I(o) > 4(1+2^{-\alpha/2})+8\sum_{k=2}^\infty \bar D_k^{-\alpha} \]
by Theorem \ref{thm:lower}. Replacing $\bar D_k$ by an upper bound $ck>\bar D_k$ still provides a lower bound.
A simple bound is $\bar D_k<k(1+\sqrt 2)/2$. 
This is an upper bound since
\[ \frac{k(1+\sqrt 2)}{2} \geqslant \frac12\left(\sqrt{k^2+i^2}+  \sqrt{k^2+(k-i)^2}\right)\,,\quad \forall 0\leqslant i\leqslant k \,.\]

So choosing $c=(1+\sqrt{2})/2\approx 1.207$ yields the
bound
\[ I(o) > 4(1+2^{-\alpha/2})+\frac{8\cdot 2^\alpha (\zeta(\alpha-1)-1)}{(1+\sqrt 2)^\alpha} \,.\]
For $\alpha=4$, this is about $5.76$. The exact value is $6.03$.
Using the simple lower bound $\zeta(\alpha-1)-1>2/(2^\alpha-4)$, $\alpha>2$, 
\[ I(o) > 4(1+2^{-\alpha/2}) + \frac{2^{\alpha+4}}{(2^\alpha-4)(1+\sqrt{2})^\alpha} \,.\]
The coefficient $c$ can be sharpened. The average distance is given by
\begin{align*}
   \bar D_k&=\frac{1}{8k} \left[4k+4k\sqrt 2+8k\sum_{j=1}^{k-1} \sqrt{1+j^2/k^2}\right]\,.
 \end{align*}
 Noting that 
 \[ \sum_{j=1}^{k-1} \sqrt{1+j^2/k^2} < (k-1) \int_0^1 \sqrt{1+x^2}\,\d x \,,\]
 we obtain the bound
\begin{align*}
  \bar D_k &\leqslant \frac12\left(1+\sqrt 2+(k-1)(\sqrt 2-\log(\sqrt 2-1))\right)\,.
\end{align*}
For $k>1$, $\bar D_k\leqslant ck$ for
\[ c\triangleq \frac{\sqrt 2}{2}+\frac14(1-\log(\sqrt{2}-1)) \approx 1.1775\,,\]
which yields the sharper lower bound
\begin{equation*}
   I(o)> 4(1+2^{-\alpha/2})+\sum_{k=2}^\infty 8k (ck)^{-\alpha} =4(1+2^{-\alpha/2})+8c^{-\alpha}(\zeta(\alpha-1)-1)\,.
 \end{equation*}
 Lower bounding the zeta function using \eqref{zeta_lower} in Lemma \ref{lem:zeta} yields the closed-form bound
 \begin{equation}
   I(o)> 4(1+2^{-\alpha/2})+8c^{-\alpha}\frac{3^{\alpha-1}+2^{\alpha-1}+1}{6^{\alpha-1}-3^{\alpha-1}-2^{\alpha-1}-1}\,.
  \label{sq_lower}
 \end{equation}
 For $\alpha=4$, this is $5.84$.

\subsection{Upper interference bounds}
Let $X_{i,j}$ be a uniformly randomly distributed variable on $[i-1/2,i+1/2]\times [j-1/2,j+1/2]$.
Then for any $\mathcal{Z}\subset\Lo$, the lattice sum \eqref{square_lattice_origin} is bounded by
\begin{align*}
  I(o)=\sum_{(i,j)\in\Z^2\setminus\{o\}} \E(X_{i,j})^{-\alpha} \leqslant &
      \sum_{(i,j)\in\mathcal{Z}} \E(X_{i,j})^{-\alpha}+\sum_{(i,j)\in\mathcal{Z}^c} \E(X_{i,j}^{-\alpha})\\
   =& \sum_{(i,j)\in\mathcal{Z}} \E(X_{i,j})^{-\alpha}+
  \int_A \|x\|^{-\alpha} \d x\,,
\end{align*}
where $\mathcal{Z}^c= \Lo\setminus\mathcal{Z}$ and 
\begin{align*}
   A&=\R^2\setminus \left( \bigcup_{(i,j)\in\mathcal{Z}} \{ [i-1/2,i+1/2]\times [j-1/2,j+1/2] \}\right) \\
    &=\mathcal{Z}^c \oplus u_\square\,,
 \end{align*}
 where $\oplus$ is the morphological dilation operator and the structuring element is the
unit square centered at $o$, \ie,
$u_\square=V_\L(o)=\{x\in\R^2\colon \|x\|_{\infty} \leqslant 1/2\}$. The bound is tighter if $\mathcal{Z}$ includes
the points closer to the origin.
For example $\mathcal{Z}=\{ x\in\Z^2\colon 0<\|x\|\leqslant \sqrt 2\}$. This gives the bound
\[ I(o) < 4(1+2^{-\alpha/2})+\int_{\R^2\setminus [0,3/2]^2} \|x\|^{-\alpha}\d x \,.\]
This integral does not have a solution for general $\alpha$. For rational $\alpha$, it can be
expressed using hypergeometric functions, and for integer $\alpha$, it simplifies to
\[ \alpha=3: \; 8\sqrt{2}/3\,;\quad \alpha=4: \; 2(2+\pi)/9\,;\quad \alpha=5: \; 80\sqrt{2}/243 \,.\]
So for $\alpha=4$, we get $I(o)<6.14$, which is within $1.5\%$ of the exact value.

Changing to polar coordinates yields simpler closed-form bounds. As an application 
of Corollary \ref{cor:radial}, we
outer-bound the area $A$ by $\R^2\setminus b_o(b)$ with $b=3/2$.
This gives
\begin{equation}
 I(o)< 4(1+2^{-\alpha/2})+\frac{2\pi (3/2)^{2-\alpha}}{\alpha-2} \,,
 \label{sq_upper}
\end{equation}
which is a decent bound for $\alpha>3$. For $\alpha=4$, $I(o)<6.40$.
Better bounds can be obtained by using a larger radius $r_b$ in Cor.~\ref{cor:radial}.
Adding 12 more nodes in the direct calculation and increasing $r_b=3/\sqrt 2$, we obtain
\begin{equation}
 I(o)<4(1+2^{-\alpha/2}+2^{-\alpha}+2\cdot 5^{-\alpha/2}) \frac{2\pi(3/\sqrt 2)^{2-\alpha}}{\alpha-2} \,.
 \label{sq_upper2}
\end{equation}
This situation is illustrated in \figref{fig:voronoi_sq}, where the circle has radius $r_b$.
The lower bound \eqref{sq_lower} and the tighter upper bound \eqref{sq_upper2} are shown in \figref{fig:square_interf}.
 
\figs
\begin{figure}
\centerline{\epsfig{file=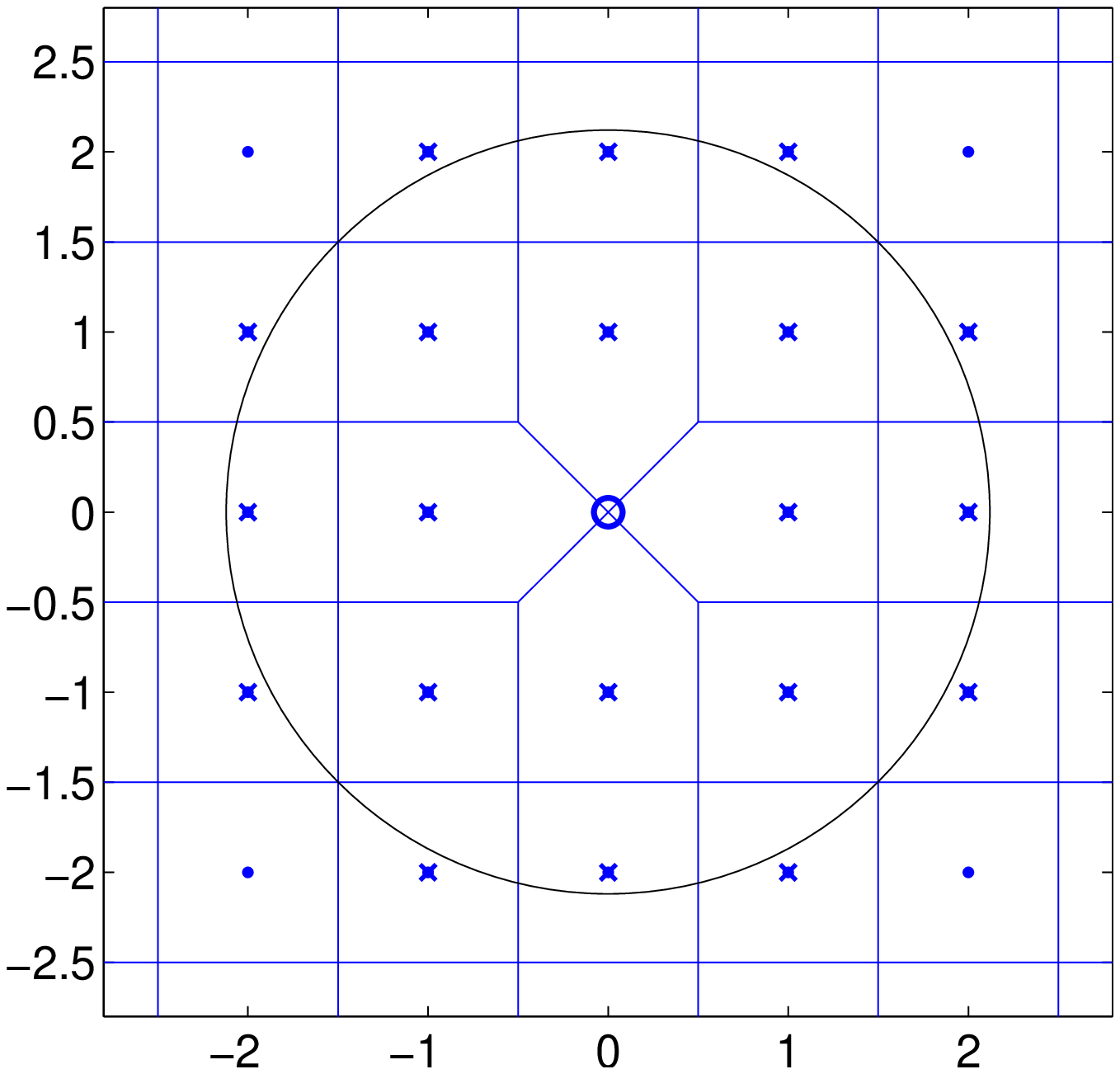,width=9cm}}
\caption{Voronoi cells for square lattice. The circle of radius $3/\sqrt{2}$ indicates the radial bound for the
integration for the bound \eqref{sq_upper2}. The interference from the 20 nearest points, marked by a cross, is summed up
directly, while the interference from the other nodes is upper bounded by integrating $\|x\|^{-\alpha}$ over the outside of the circle.}
\label{fig:voronoi_sq}
\end{figure}

\begin{figure}
\centerline{\epsfig{file=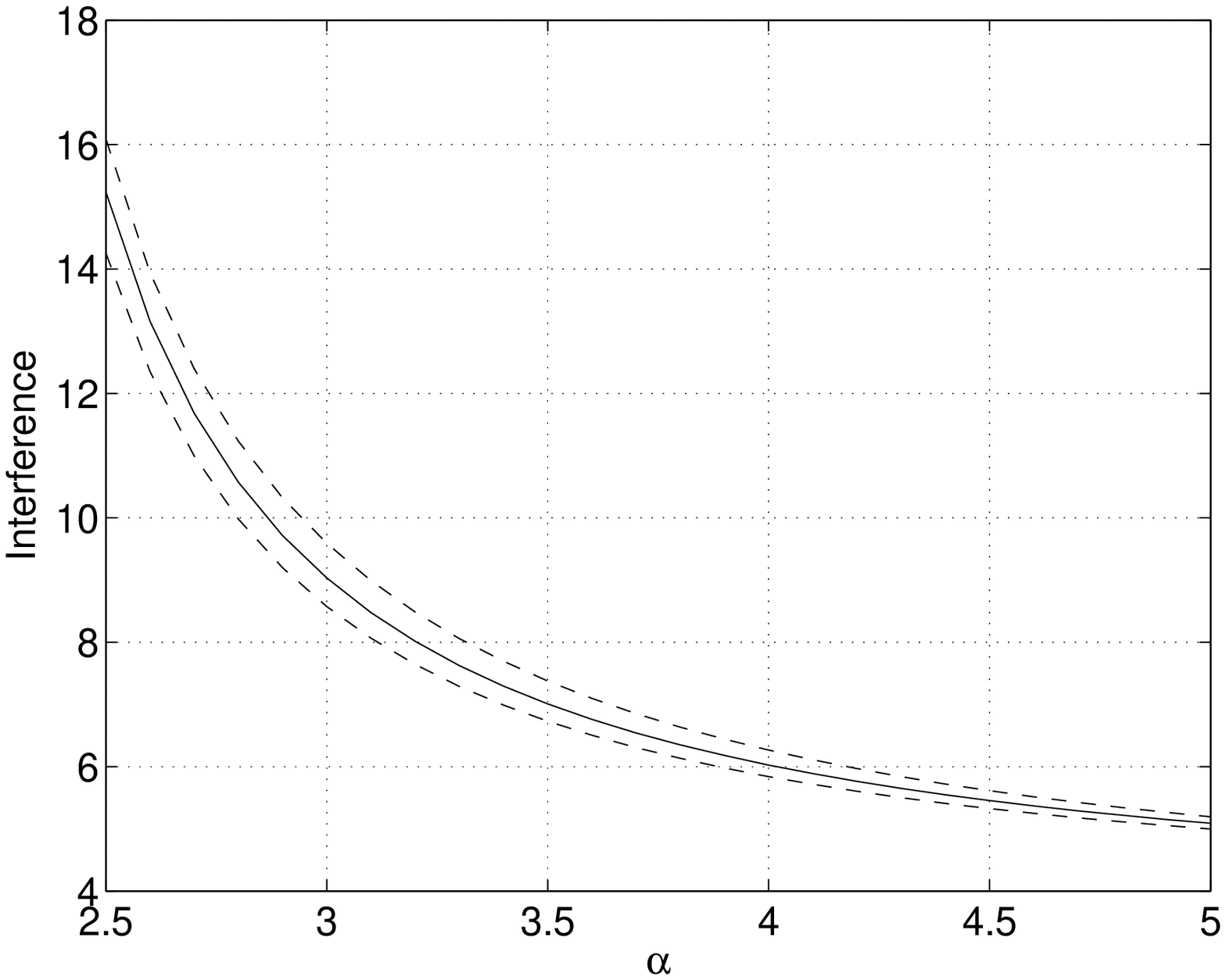,width=9cm}}
\caption{Interference at the origin in a square lattice (solid curve) and lower bound \eqref{sq_lower} and upper bound
\eqref{sq_upper2} (dashed). }
\label{fig:square_interf}
\end{figure}
\fi

\subsection{Transmitter-receiver offset}
Let $I(z)$ be the interference measured at location $z$, with $\|z\|<1$.
First we consider the case where $z$ lies on the positive real axis, \ie, $z=(\|z\|,0)$. Let $r=\|z\|$.
Let $I_8$ denote the interference from the $8$ nearest interferers.
From elementary geometry, the distance $x$ to interferer $(1,1)$ is given by $x^2=2-r(2-r)$,
and the distance $x'$ to $(-1,-1)$ is given by $x'^2=2+r(2+r)$. Hence 
\begin{align*}
   I_8(r)&=(1-r)^{-\alpha}+(1+r)^{-\alpha}+2(2-r(2-r))^{-\alpha/2}+2(2+r(2+r))^{-\alpha/2}+2(1+r^2)^{-\alpha/2} \\
   &=\underbrace{4(1+2^{-\alpha/2})}_{I_8(o)}+\underbrace{\alpha^2(1+2^{-\alpha/2-1})}_{c_8} r^2+O(r^4) \,.
\end{align*}
A simple lower bound for $2<\alpha<12$ on the excess interference coefficient is $c_8>\alpha^2+1$. 
Including the interference from all other nodes:
\[ I(z)  <  I_8(r)+ \int_{\R_+^2\setminus [0,3/2]^2 }\|x-z\|^{-\alpha}\d x \]
For $\alpha=3$,
\[ I(z)\approx 4(1+2^{-3/2})+\underbrace{\left(9(1+2^{-5/2})+\frac{20\sqrt 2}{27}+\frac{32}{27}\right)}_{\c} \|z\|^2 \,,\]
and for $\alpha=4$,
\[ I(z)\approx 5+\underbrace{\left(18+\frac{4\pi}{9}+\frac{32}{81}\right)}_{\c} \|z\|^2 \,. \]
Next, consider the case where the receiver is displaced diagonally, \ie, $z=(r,r)/\sqrt 2$.
In this case, we have
\begin{align*}
   I'_8(r)&=(\sqrt2-r)^{-\alpha}+(\sqrt2+r)^{-\alpha}+2(1+r(r-\sqrt2))^{-\alpha/2}+2(1+r(\sqrt2+r))^{-\alpha/2}+2(2+r^2)^{-\alpha/2} \\
   &=4(1+2^{-\alpha/2})+\underbrace{\alpha^2(1+2^{-\alpha/2-1})}_{c_8'} r^2+O(r^4) \,.
\end{align*}
Since $c_8=c_8'$, it turns out that the excess interference coefficient is the same as for the case of axial displacement.

\subsection{Application to TDMA}
\subsubsection{Separating the transmitters}
If the transmitting nodes are located at $(m\Z)^2$, corresponding to an $m^2$-phase TDMA scheme, the interference at
the receiver $(0,1)$ is $I_m=m^{-\alpha} I(1/m)$, as in the one-dimensional case. For $m>2$, the nearest 8 interferers are at the following
distances: One at $m-1$, two at $\sqrt{1+m^2}$, two at $\sqrt{(m-1)^2+m^2}$, one at $m+1$, and two at $\sqrt{(m+1)^2+m^2}$. 
If these 8 terms are summed up directly and the Voronoi upper bound in Cor.~\ref{cor:radial} is used to upper bound the interference
from the other nodes, the radius $r_b=3m/2-1$, since the 9-th nearest node is at distance $2m-1$ and the Voronoi cell is $m/2$ wide.

\subsubsection{Balanced TDMA}
As in the one-dimensional case, there is a smarter way to schedule the transmissions, such that the two nearest
interferers are at the same distance. For all $m>1$, there exists a scheme such that the eight nearest interferers are:
Four at distance $\sqrt{m^2+1}$, two at $\sqrt{m^2+(m-1)^2}$, and two at $\sqrt{m^2+(m-1)^2}$. The smallest distance from the
receiver to a Voronoi cell of an interferer not included in these eight is $3m/2$.

\figs
\begin{figure}
\centerline{\epsfig{file=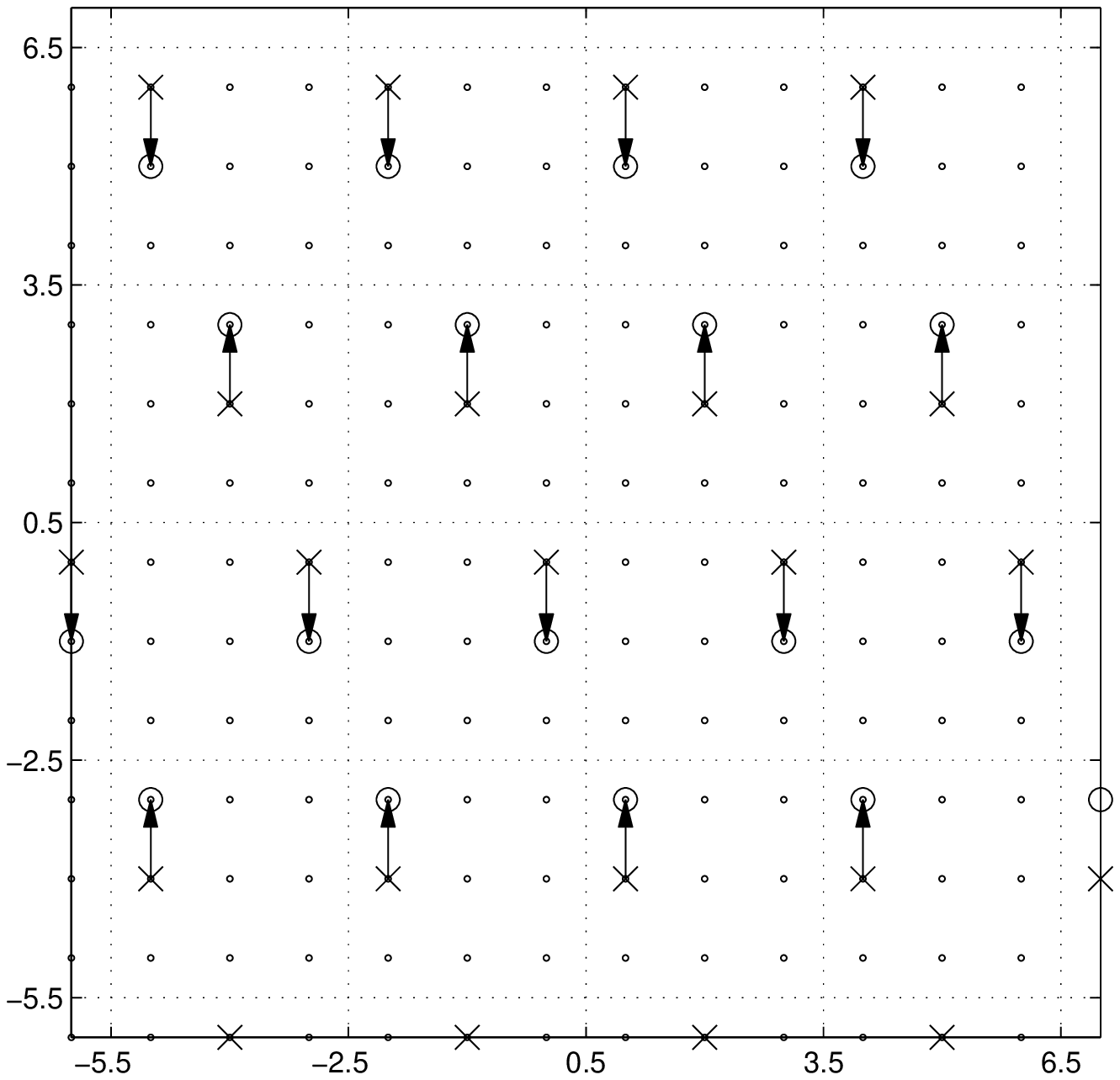,width=9cm}}
\caption{Balanced TDMA scheme for square lattice for $m=3$. The small dots are the lattice points $\Z^2$, the crosses
the transmitters, and the circles their receivers. Arrows indicate transmissions. The dotted lines demarcate
the $3\times 3$ boxes, in which there is one transmitter in each time slot. The transmit-receive pattern can be shifted
and rotated, so that in $4\cdot 3^2$ time slots, each node gets an opportunity to transmit to all four nearest neighbors
with the same interference.}
\label{fig:sq_balanced}
\end{figure}
\fi

We believe this balanced scheme is optimum in the sense that for all sub-lattices in which one node per
$m\times m$ block transmits in each slot, this one
causes the smallest interference. From the triangular lattice it is known that we cannot do better than having the six nearest
interferers at distance $\sqrt{2/\sqrt 3}m$, and the average distance of the nearest six neighbors lies within 3\%-5\% of this bound for
all $m\leq 30$. \figref{fig:sq_tdma} illustrates the interference gain of the balanced scheme when compared with the simple one.

\figs
\begin{figure}
\centerline{\epsfig{file=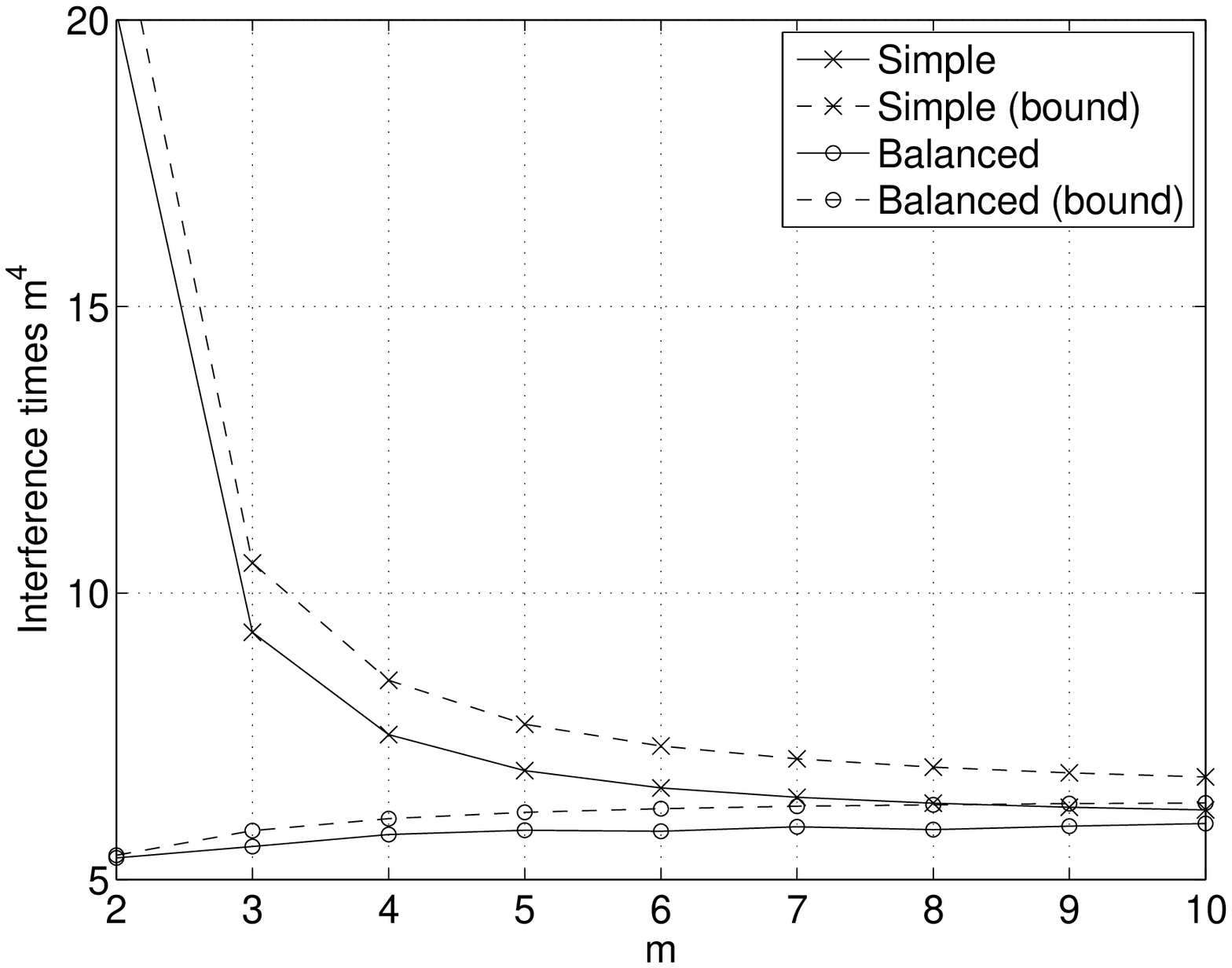,width=9cm}}
\caption{Interference for TDMA scheme for square lattices with $\alpha=4$, multiplied by $m^4$ for normalization.
The ``simple" scheme is the one where nodes $(m\Z)^2$ transmit, the ``balanced" one is the one illustrated in \figref{fig:sq_balanced}.
The dashed curves are the bounds obtained by the radial bounds in Cor.~\ref{cor:radial}. As $m\to\infty$, the curves converge
to the same value.}\label{fig:sq_tdma}
\end{figure}
\fi

\section{Triangular lattices}
A triangular lattice, where nodes are arranged in a lattice with generator matrix $\mathbf{G}_\mathrm{tri}$ given in
\eqref{generators}, offers the densest packing given the nearest-neighbor distance. It is the preferred deployment
for sensor networks with isotropic sensors, since the smallest number of sensors is needed to cover an area.
 The density is $2/\sqrt{3}$.
More importantly though,
a triangular lattice may be a good model for a CSMA-type network. Assume a high-density network, with $\lambda\gg 1$ nodes
per unit area, and a CSMA scheme with carrier sensing radius $1$. With ideal sensing, the transmitting nodes cannot be
denser than a triangular grid. Hence the interference in a triangular lattice is an upper bound to the interference in a
CSMA network if transmitters are spaced at least at unit distance.

\subsection{Lower interference bound}
Again we can partition the interferers into rings of increasing radii; in this case, the rings are hexagons.
The average distance to the interferers in the $k$-th ring
is $k(1/2+\sqrt 3/4)\approx 0.933k$, which, using Theorem \ref{thm:lower}, immediately yields the lower bound
\[ I(o) > 6+6\sum_{k=2}^\infty k \left[k\left(\frac12+\frac{\sqrt 3}{4}\right)\right]^{-\alpha}=6+6\left(\frac4{2+{\sqrt 3}}\right)^{\alpha}
(\zeta(\alpha-1)-1)\,. \]
Lower bounding $\zeta(\alpha)$ using Lemma \ref{lem:zeta}, we obtain
\begin{equation}
   I(o) > 6+\left(\frac4{2+{\sqrt 3}}\right)^{\alpha} \frac{2\cdot 3^\alpha+3\cdot 2^\alpha+6}{6^{\alpha-1}-3^{\alpha-1}-2^{\alpha-1}-1} \,.
   \label{tri_lower}
\end{equation}

\subsection{Upper interference bounds}
Here we apply Cor.~\ref{cor:radial} with $r_b=2/\sqrt 3$, which is the smallest distance to a Voronoi cell of a point at distance 
$\sqrt{3}$ from the origin. This implies that the six nearest neighbors at distance $1$ are considered separately, and
all other points are included in the approximation. The area of a Voronoi cell is $V=\sqrt 3/2$, hence the bound is
\begin{equation}
 I(o)< 6+\frac{4\pi}{\sqrt 3} \frac{(2/\sqrt 3)^{2-\alpha}}{\alpha-2} \,.
\end{equation}
A tighter bound is obtained by taking the nearest 18 nodes (two rings) and integrating over the Voronoi cells of the other points.
There are six nodes at distance $\sqrt 3$ and six at distance 2, and $r_b=\sqrt{4+1/3}$. This situation is illustrated in
\figref{fig:voronoi}, where the circle has radius $r_b$.
It follows that
\begin{equation}
I(o)<6(1+2^{-\alpha}+3^{-\alpha/2})+\frac{4\pi}{\sqrt 3} \frac{(13/3)^{1-\alpha/2}}{\alpha-2}\,.
\label{tri_upper2}
\end{equation}
The lower bound \eqref{tri_lower} and the tighter upper bound \eqref{tri_upper2} are shown in \figref{fig:triang_interf}.

\figs
\begin{figure}
\centerline{\epsfig{file=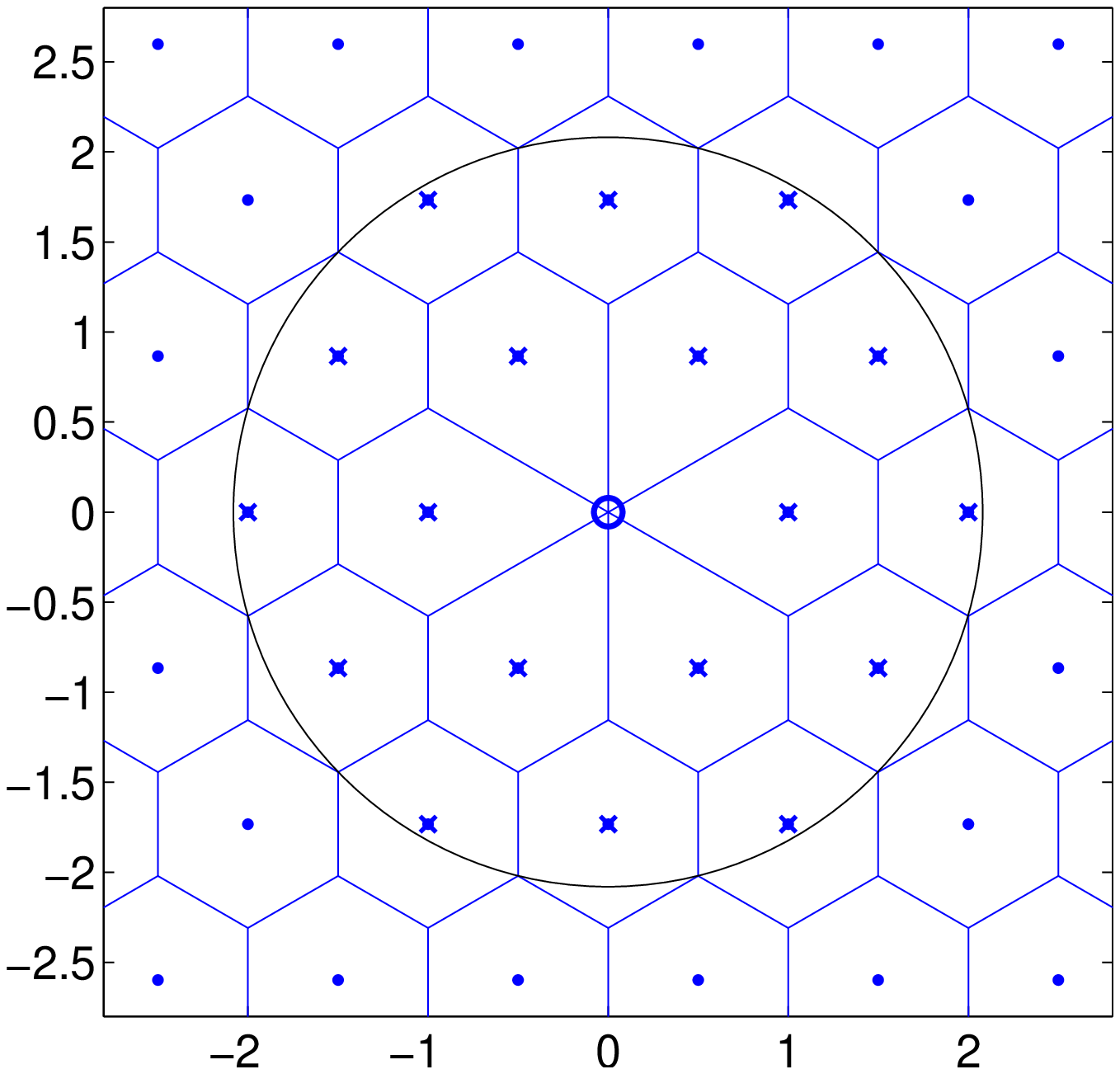,width=9cm}}
\caption{Voronoi cells for triangular lattice. The circle of radius $\sqrt{13/3}$ indicates the radial bound for the
integration for the bound \eqref{tri_upper2}. The interference from the 18 nearest points, marked by a cross, is summed up
directly, while the interference from the other nodes is upper bounded by integrating $\|x\|^{-\alpha}$ over the outside of 
the circle
and multiplying by the lattice density $1/V=2/\sqrt 3$. }
\label{fig:voronoi}
\end{figure}

\begin{figure}
\centerline{\epsfig{file=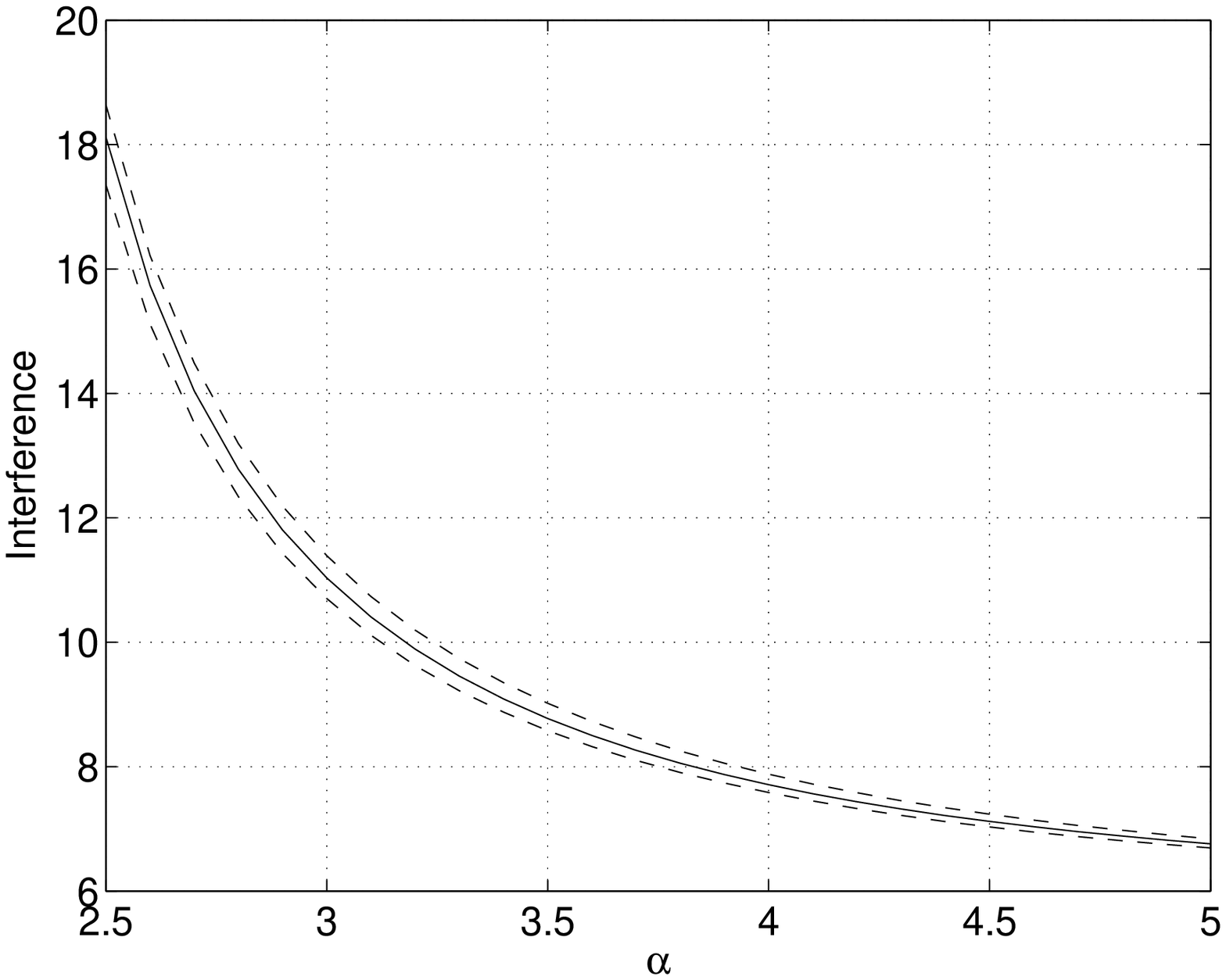,width=9cm}}
\caption{Interference at the origin in a triangular network (solid) with lower bound \eqref{tri_lower} and upper bound \eqref{tri_upper2}
(dashed). The two bounds are uniformly tight.}
\label{fig:triang_interf}
\end{figure}
\fi

\subsection{Transmitter-receiver offset}
Here we approximate the interference $I(z)$ for $\|z\|<1$. As for the square lattice, we focus first on the case where
$z$ lies on the line between the origin and the node at $(1,0)$, \ie, $z=(\|z\|,0)$, and we let $r=\|z\|$.
The interference from the 6 nearest interferers is given by 
\[ I_6(r)=(1-r)^{-\alpha}+(1+r)^{-\alpha}+2(1-r(1-r))^{-\alpha/2}+2(1+r(1+r))^{-\alpha/2} \]
since the distance $x$ of point $(r,0)$ to interferer $(1/2,\sqrt{3}/2)$ is given by $x^2=1-r(1-r)$.
A Taylor expansion gives
\[ I_6(r)=6+\frac32\alpha^2 r^2+O(r^4)\,. \]
Taking a lower bound on the interference from the other nodes yields a lower bound for the total interference:
\begin{equation}
   I(r)> \underline{I}(o)+\frac32\alpha^2 r^2\,, 
   \label{tri_offset}
\end{equation}
where $\underline{I}(o)$ denotes the lower bond on the right-hand side of \eqref{tri_lower}.
A better approximation may be obtained by decreasing the radius $r_b$ used for the upper bounds by $r$,
as stated in Cor. \ref{cor:radial}, but the effect of a small displacement $r$ on this term is relatively small.
This is illustrated in \figref{fig:triang_offset}, where the actual interference and the quadratic lower bound \eqref{tri_offset}
are shown. The curvature of the interference $I(r)$ near $r=0$, \ie, the excess interference, is reproduced quite accurately
by the bound. It can further be observed that
while higher path loss exponents result in smaller interference for no or very small displacements, there is a cross-over
point after which a {\em smaller} $\alpha$ is better.

Changing the direction of the displacement $z$ does not affect the quadratic approximation, \ie, the difference only
manifests itself for $r>1/4$. So in the triangular case, the offset coefficient $\c\approx 3\alpha^2/2$ for all directions.

\figs
\begin{figure}
\centerline{\epsfig{file=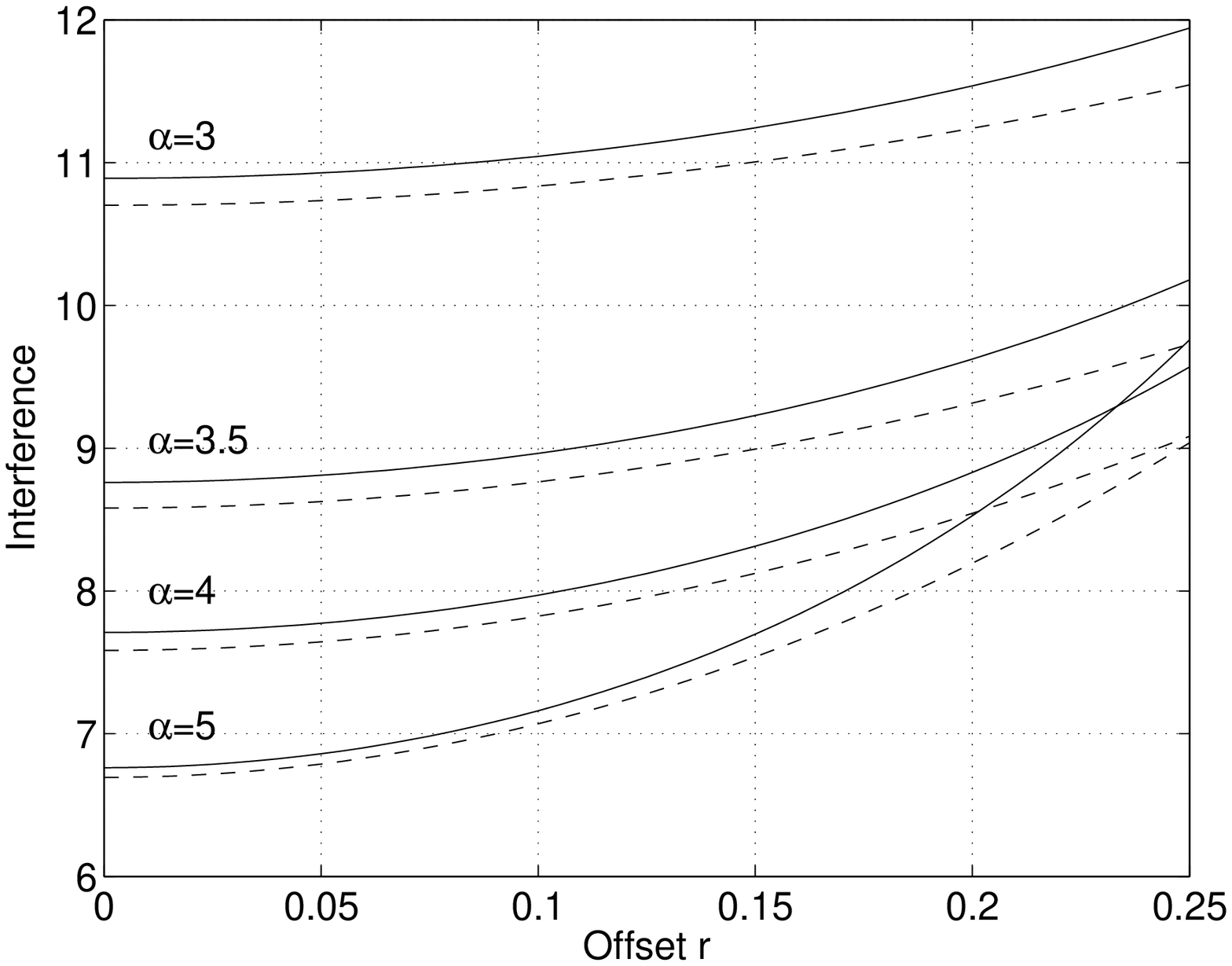,width=9cm}}
\caption{Interference at position $(r,0)$ in a triangular network for $\alpha=3,3.5,4,5$. The solid curves are the exact numerical results,
while the dashed ones are the bounds \eqref{tri_offset}.}
\label{fig:triang_offset}
\end{figure}
\fi

\subsection{Application to TDMA}
\subsubsection{Separating the transmitters}
The most straightforward TDMA scheduling scheme for the triangular lattice is probably the one where one node in a rhombus
containing $m^2$ nodes transmits. The generator matrix for the transmitting lattice is then
\[ \mathbf{G}_\mathrm{rh}=\begin{bmatrix} m& -\mathbf{1}_{\{m \text{ odd}\}}/2 \\0 & m\sqrt 3/2 \end{bmatrix}\,. \]
For even $m$, the transmitters form a rectangular lattice with horizontal spacing $m$ and vertical spacing $m\sqrt 3/2$.
If the receiver is to the right of each transmitter, the interference is acceptable, but this is certainly not the optimum scheme,
since the nearest interferer is at distance $m-1$.
An improved scheduler lets one node in a parallelogram containing $(m+1)m$ nodes transmit:
\[ \mathbf{G}_\mathrm{par}=\begin{bmatrix} m+1& -m/2 \\0 & m\sqrt 3/2 \end{bmatrix}\, \]
\subsubsection{Balanced TDMA}
An even better scheme is the one inspired by the balanced one-dimensional scheme, where in each horizontal
row of nodes, the pattern $(\overrightarrow{\text{T}}$R$^{m-1})_m$ is employed.
This way, the four closest interferers are all at distance $m$, and the next two at $\sqrt{(m/2+1)^2+3m^2/4}$. Since
these distances are close to the sphere-packing bound, this may be the optimum scheme. 

The three TDMA MACs
are compared in \figref{fig:tri_tdma}.

\figs
\begin{figure}
\centerline{\epsfig{file=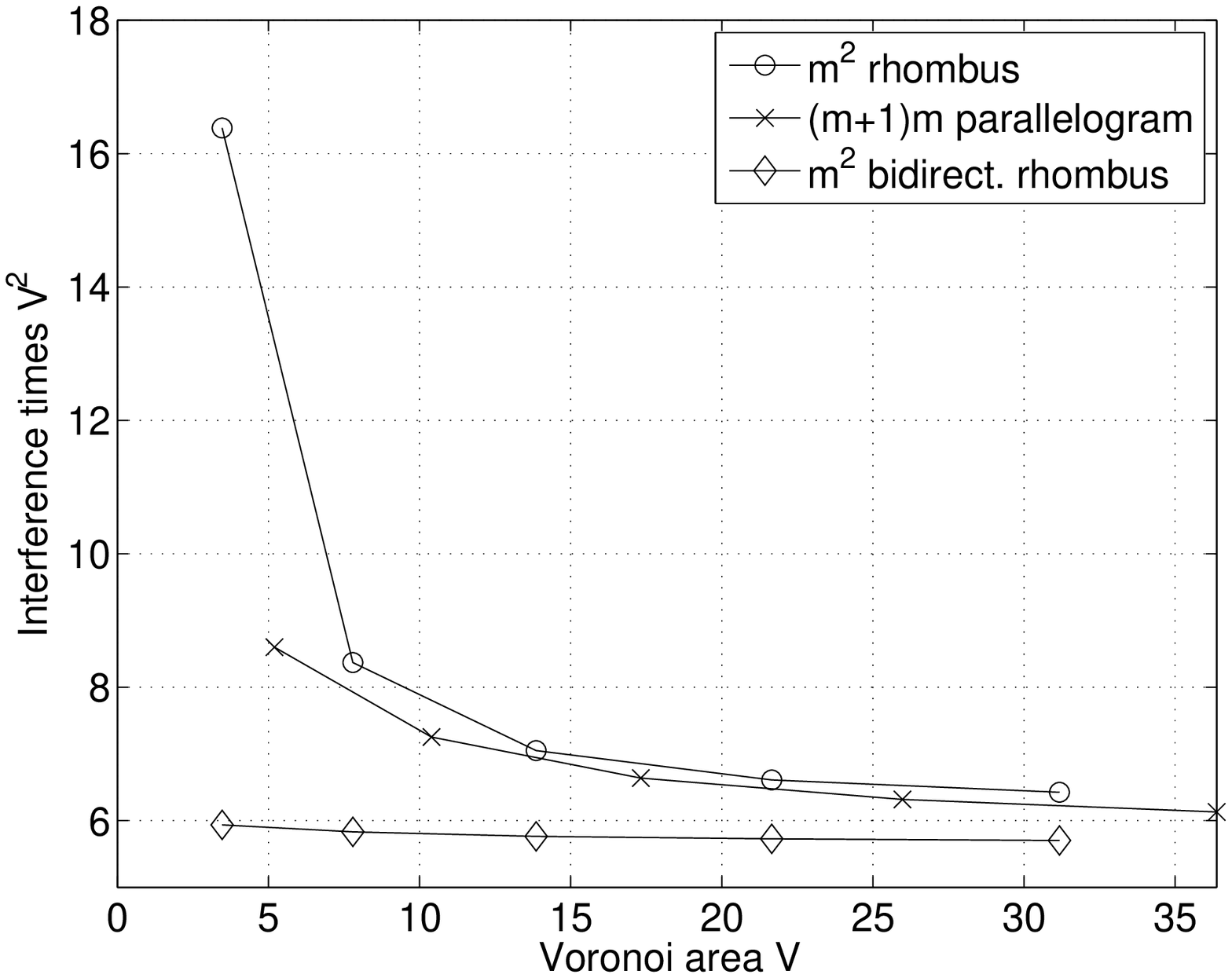,width=9cm}}
\caption{Interference for TDMA schedulers in the triangular lattice for $\alpha=4$.
For a fair comparison, the area of the corresponding Voronoi cells is used as the $x$ axis, and the $y$ axis is 
scaled by $\lambda^{-\alpha/2}=\lambda^{-2}$, where $\lambda=1/V$, since scaling the network by a factor $s$
in both dimensions scales the density by $s^{-2}$ and the interference by $s^{-\alpha}$.}
\label{fig:tri_tdma}
\end{figure}
\fi

\section{Conclusions}
By judicious application of Jensen's inequality, tight and general upper and lower bounds on the interference in
lattice networks can be calculated. The lower bounding technique yields expressions involving the Riemann zeta function,
which, in turn, can be tightly lower bounded in closed-form (Lemma \ref{lem:zeta}).

When transmitters are arranged in a lattice, but their receivers are not, such as in a cellular downlink scenario or when
a MAC scheme focuses on separating the transmitters, the interference needs to be characterized at some distance $r$
from the desired transmitter. We have introduced a simple yet effective quadratic approximation that shows how the
interference increases with growing link distance $r$. In terms of the signal-to-interference-and-noise ratio, increasing $r$
has two negative effects: it reduces the strength of the desired signal, and it increases the interference. Our framework captures both.

For the case where all nodes in a network form a one-dimensional, square, or triangular network, we have analyzed the
interference induced by basic TDMA schemes, and we have suggested superior {\em balanced} schemes that are
optimum or near-optimum. It turns out that scheduling transmitter-receiver pairs instead of just transmitters significantly enhances
the throughput.

The interference bounds are also useful to avoid long simulation times. Normally as $\alpha\downarrow 2$ (in two-dimensional networks), the diameter of the simulated networks has to be impractically large just to get a good approximation of the total interference. The upper bounds obtained from integration over Voronoi cells (Theorem \ref{thm:voronoi}) demonstrate that it is possible, with negligible loss in accuracy, to assume that the interference from nodes outside a certain radius is the same as the {\em expected} interference in a network where these nodes are distributed as a homogeneous Poisson point process.

\bibliographystyle{IEEEtr}

\figs
\else
\clearpage
{\Large\bf Figures}

\begin{figure}[h]
\centerline{\epsfig{file=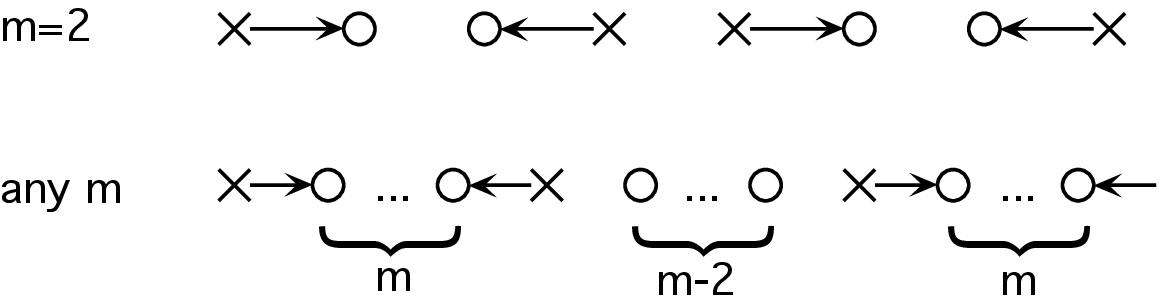,width=9cm}}
\caption{Balanced TDMA schemes for $m=2$ and general $m$. The two nearest interferers are at distances $m$ (to the left) and $m$ (to the right).}
\label{fig:smart_tdma}
\end{figure}

\begin{figure}[h]
\centerline{\epsfig{file=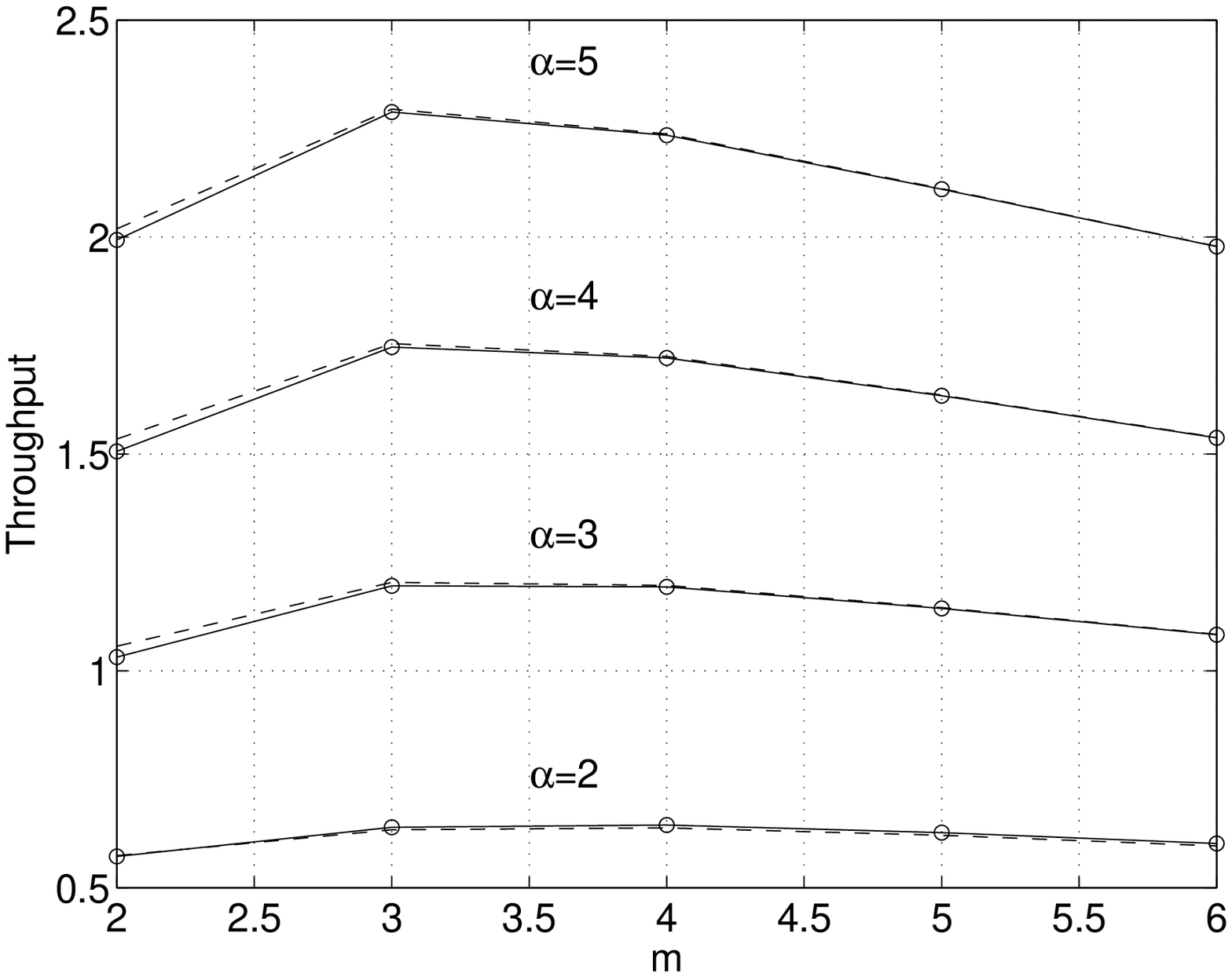,width=9cm}}
\caption{Throughput as a function of path loss exponent $\alpha$ and TDMA parameter $m$ for balanced schemes in
one-dimensional networks. The dashed lines are the approximations $\log_2(1+m^\alpha C(\alpha))/m$.}
\label{fig:onedim_through}
\end{figure}
\clearpage

\begin{figure}
\centerline{\epsfig{file=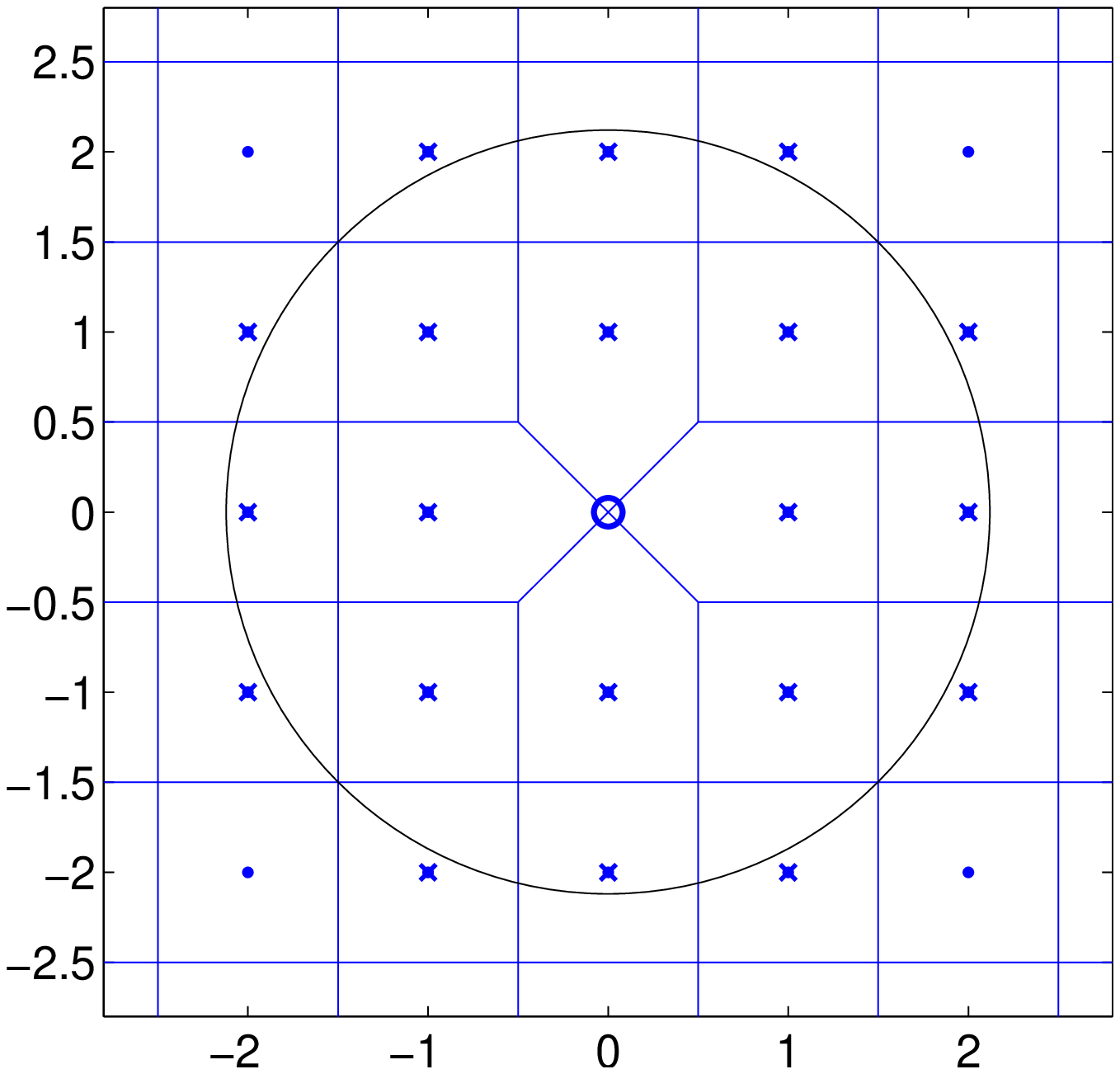,width=9cm}}
\caption{Voronoi cells for square lattice. The circle of radius $3/\sqrt{2}$ indicates the radial bound for the
integration for the bound \eqref{sq_upper2}. The interference from the 20 nearest points, marked by a cross, is summed up
directly, while the interference from the other nodes is upper bounded by integrating $\|x\|^{-\alpha}$ over the outside of the circle.}
\label{fig:voronoi_sq}
\end{figure}

\begin{figure}
\centerline{\epsfig{file=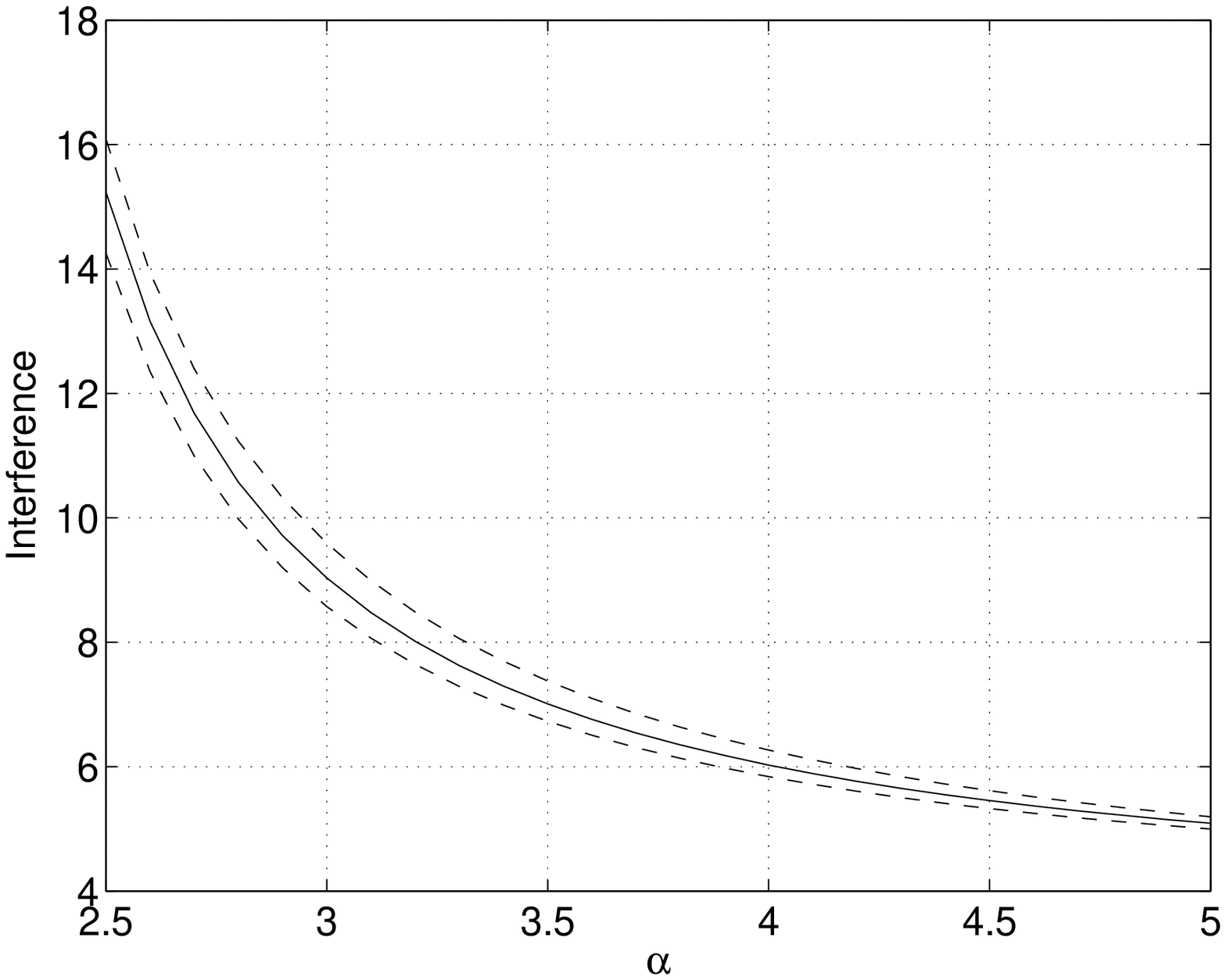,width=9cm}}
\caption{Interference at the origin in a square lattice (solid curve) and lower bound \eqref{sq_lower} and upper bound
\eqref{sq_upper2} (dashed). }
\label{fig:square_interf}
\end{figure}

\begin{figure}
\centerline{\epsfig{file=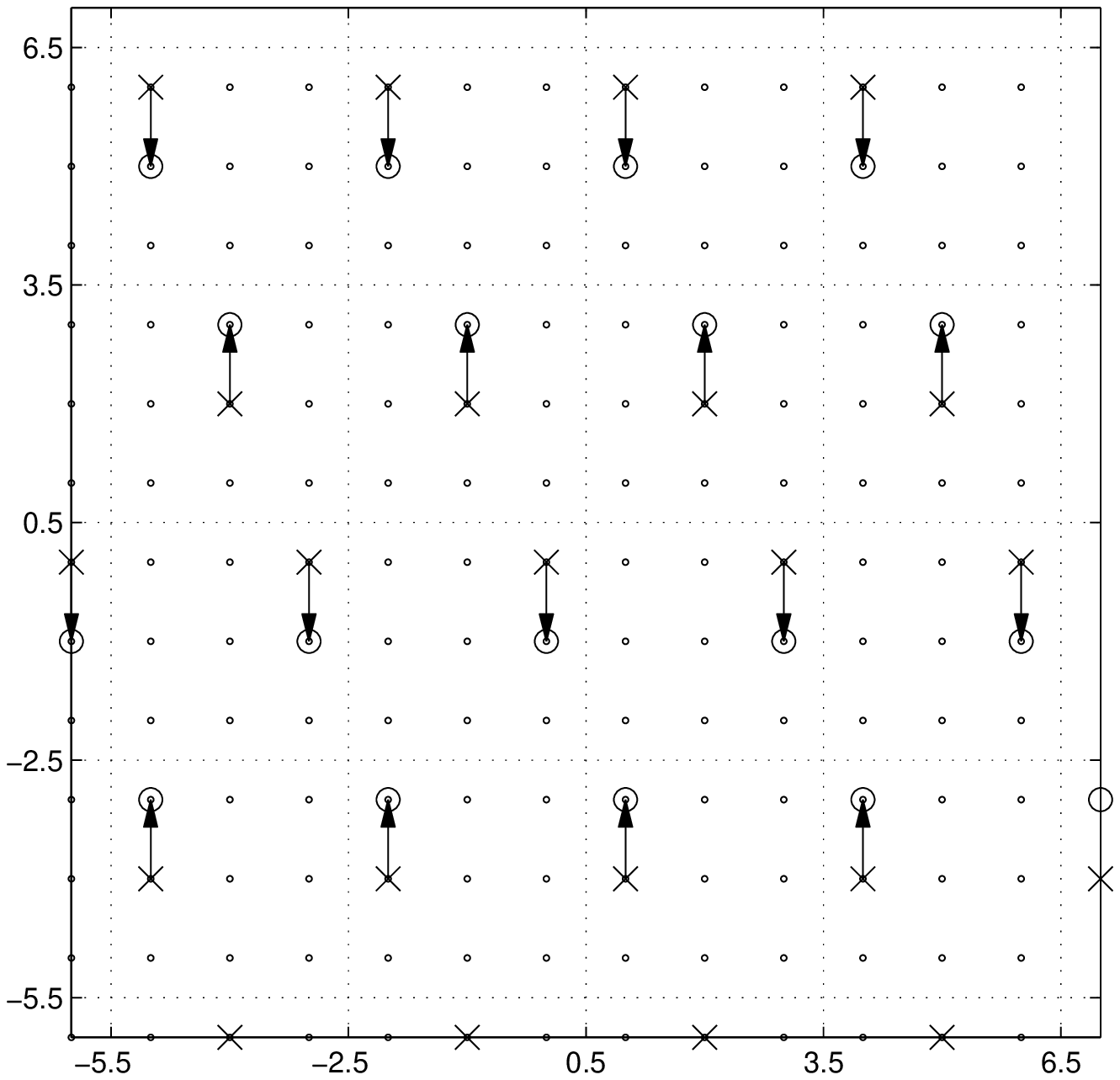,width=9cm}}
\caption{Balanced TDMA scheme for square lattice for $m=3$. The small dots are the lattice points $\Z^2$, the crosses
the transmitters, and the circles their receivers. Arrows indicate transmissions. The dotted lines demarcate
the $3\times 3$ boxes, in which there is one transmitter in each time slot. The transmit-receive pattern can be shifted
and rotated, so that in $4\cdot 3^2$ time slots, each node gets an opportunity to transmit to all four nearest neighbors
with the same interference.}
\label{fig:sq_balanced}
\end{figure}

\begin{figure}
\centerline{\epsfig{file=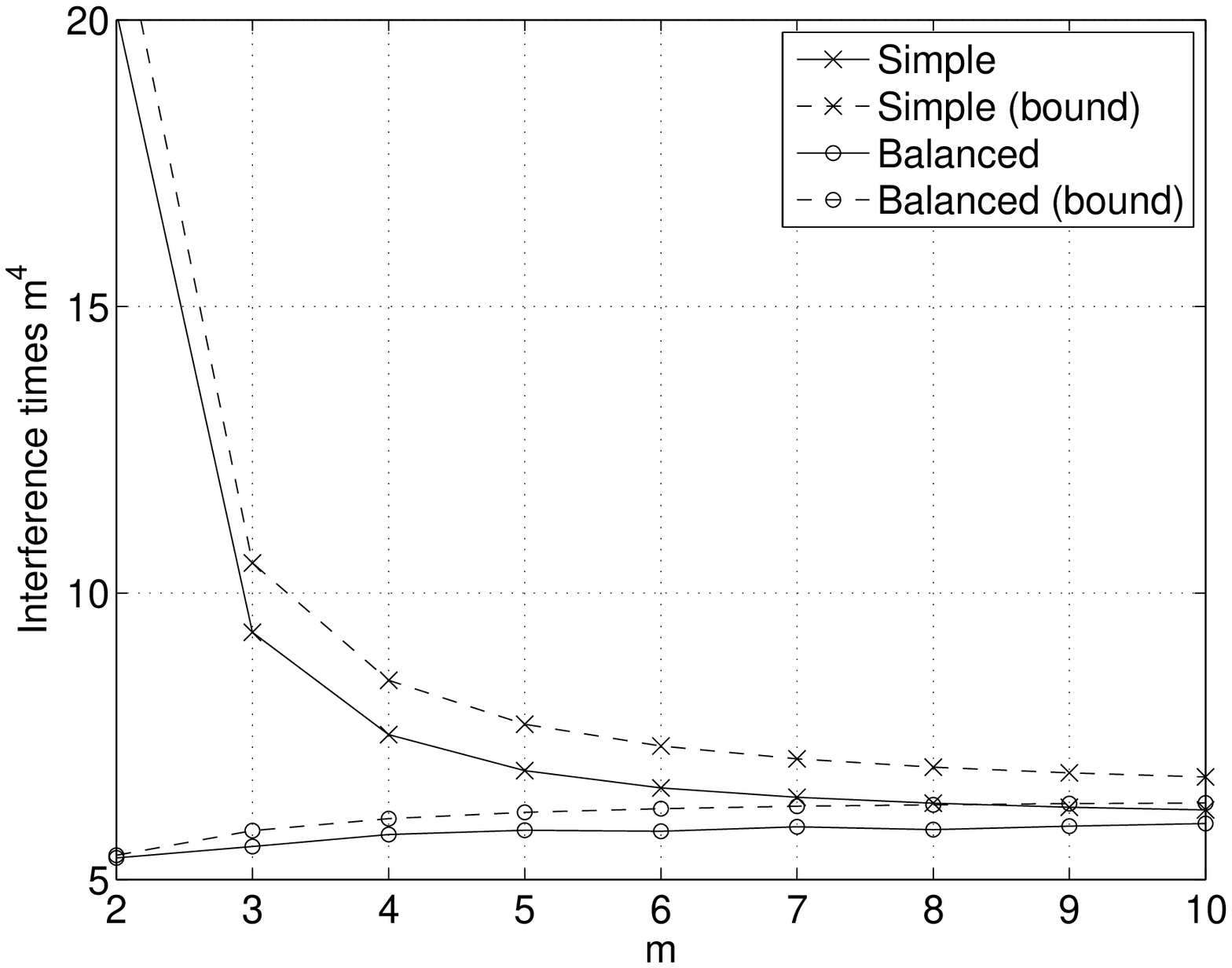,width=9cm}}
\caption{Interference for TDMA scheme for square lattices with $\alpha=4$, multiplied by $m^4$ for normalization.
The ``simple" scheme is the one where nodes $(m\Z)^2$ transmit, the ``balanced" one is the one illustrated in \figref{fig:sq_balanced}.
The dashed curves are the bounds obtained by the radial bounds in Cor.~\ref{cor:radial}. As $m\to\infty$, the curves converge
to the same value.}\label{fig:sq_tdma}
\end{figure}

\begin{figure}
\centerline{\epsfig{file=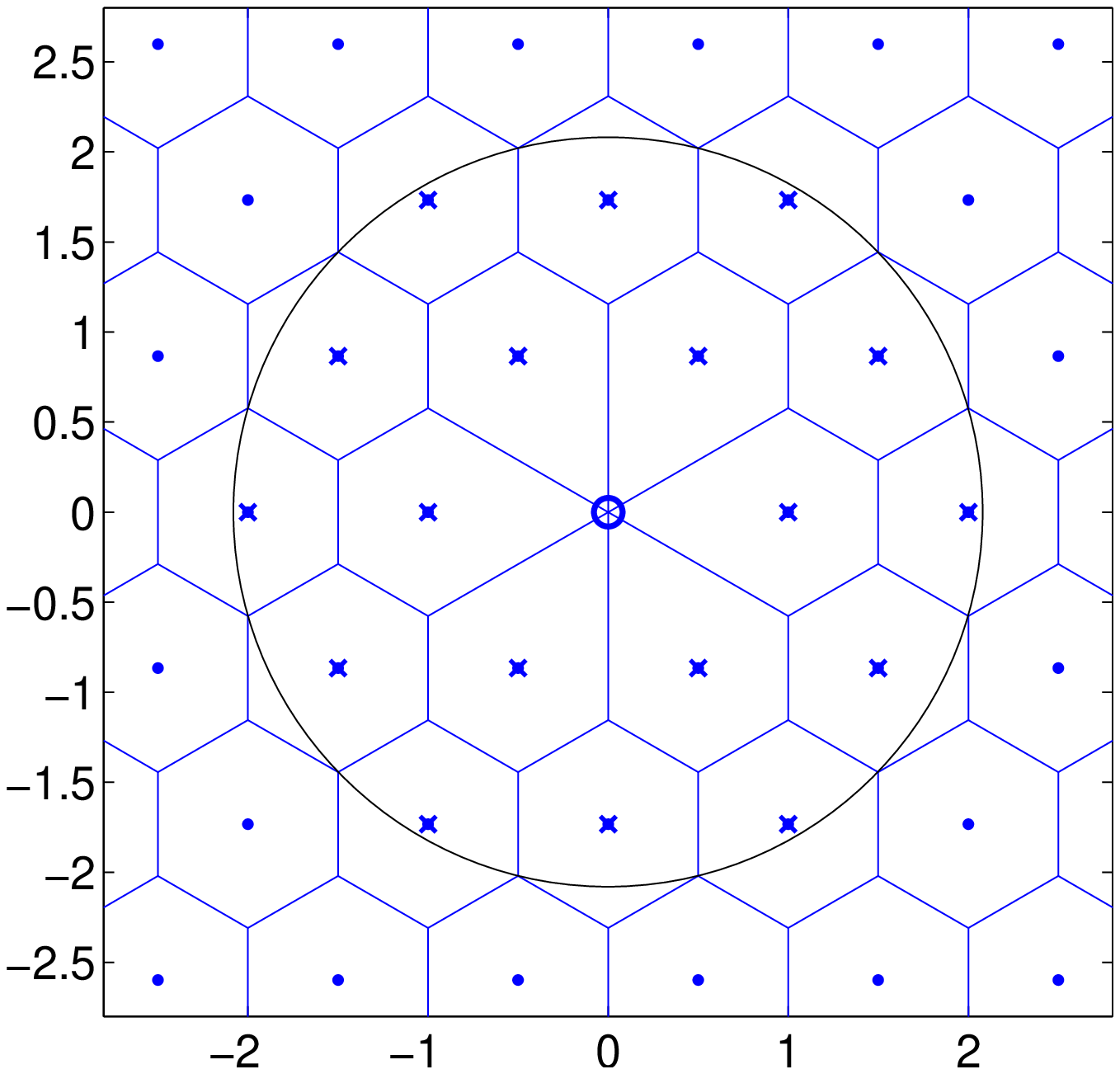,width=9cm}}
\caption{Voronoi cells for triangular lattice. The circle of radius $\sqrt{13/3}$ indicates the radial bound for the
integration for the bound \eqref{tri_upper2}. The interference from the 18 nearest points, marked by a cross, is summed up
directly, while the interference from the other nodes is upper bounded by integrating $\|x\|^{-\alpha}$ over the outside of the
circle
and multiplying by the lattice density $1/V=2/\sqrt 3$. }
\label{fig:voronoi}
\end{figure}

\begin{figure}
\centerline{\epsfig{file=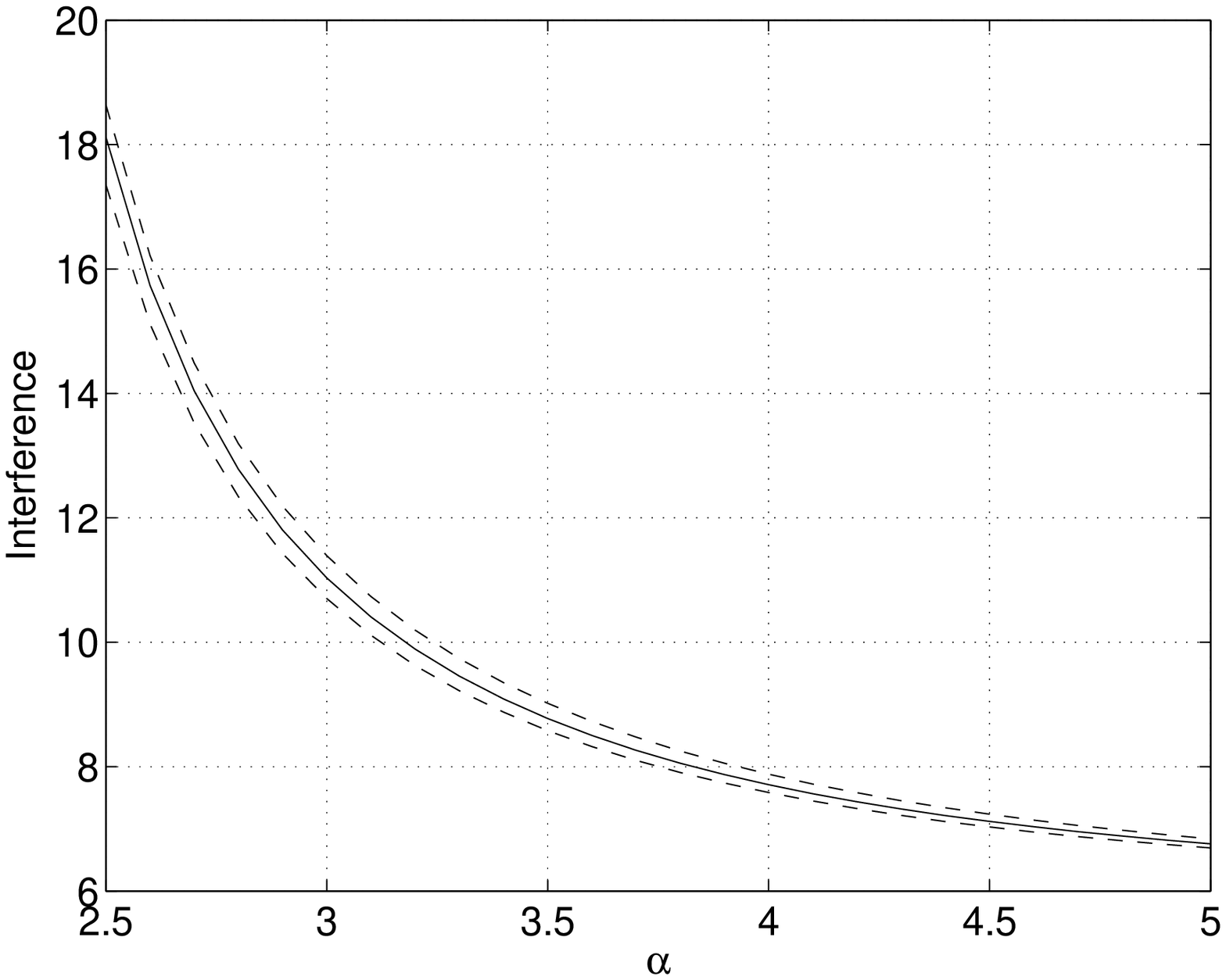,width=9cm}}
\caption{Interference at the origin in a triangular network (solid) with lower bound \eqref{tri_lower} and upper bound \eqref{tri_upper2}
(dashed). The two bounds are uniformly tight.}
\label{fig:triang_interf}
\end{figure}

\begin{figure}
\centerline{\epsfig{file=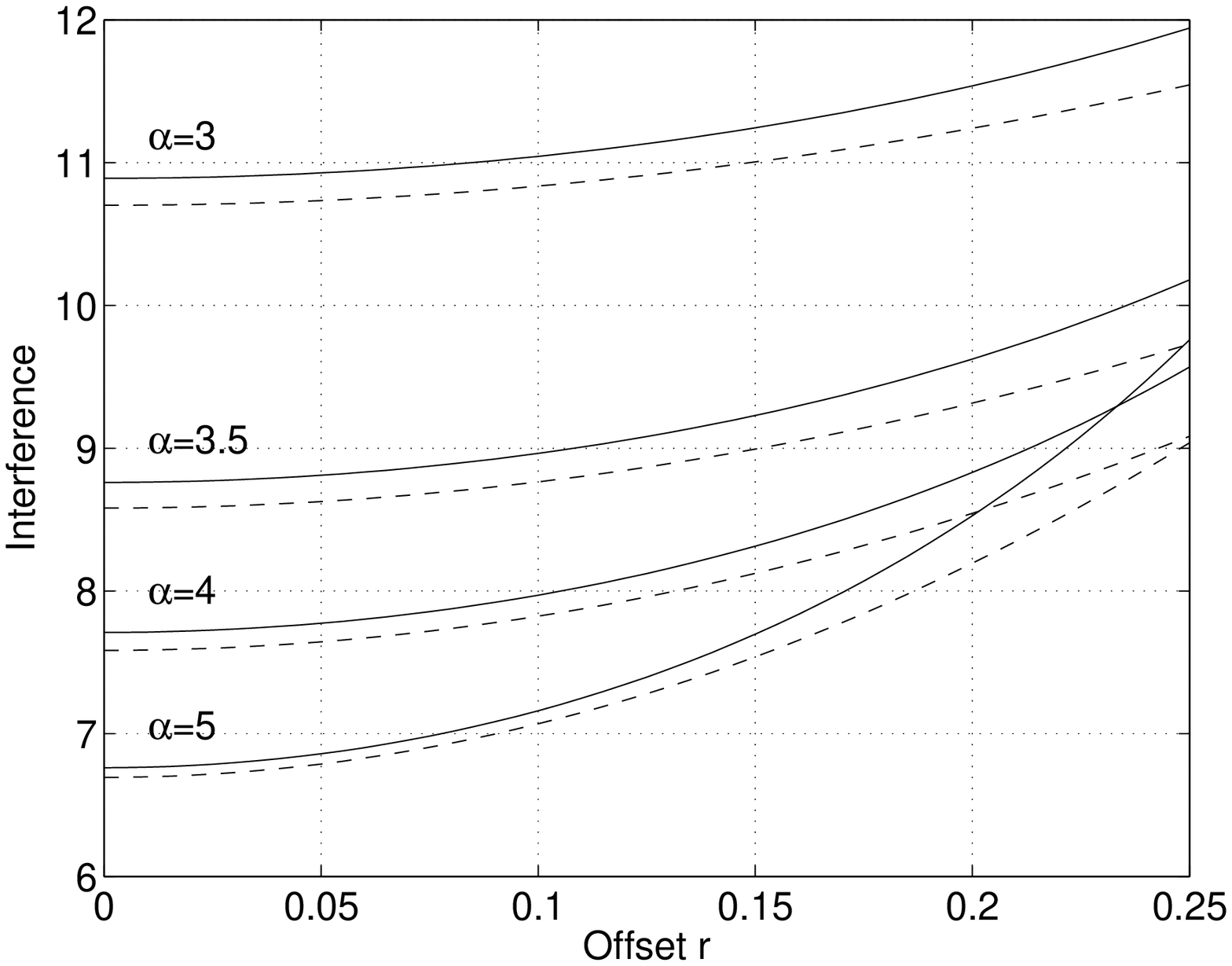,width=9cm}}
\caption{Interference at position $(r,0)$ in a triangular network for $\alpha=3,3.5,4,5$. The solid curves are the exact numerical results,
while the dashed ones are the bounds \eqref{tri_offset}.}
\label{fig:triang_offset}
\end{figure}

\begin{figure}
\centerline{\epsfig{file=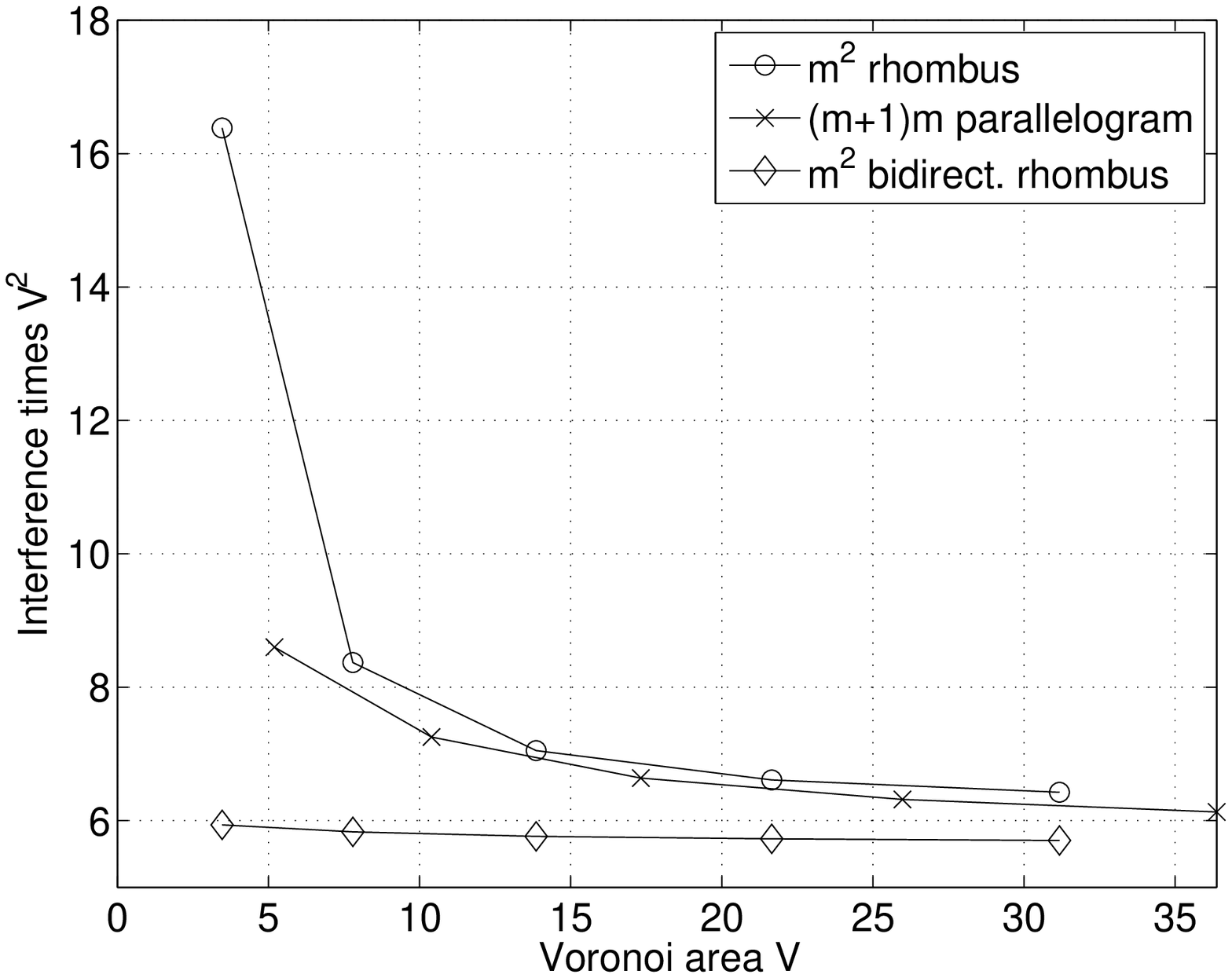,width=9cm}}
\caption{Interference for TDMA schedulers in the triangular lattice for $\alpha=4$.
For a fair comparison, the area of the corresponding Voronoi cells is used as the $x$ axis, and the $y$ axis is 
scaled by $\lambda^{-\alpha/2}=\lambda^{-2}$, where $\lambda=1/V$, since scaling the network by a factor $s$
in both dimensions scales the density by $s^{-2}$ and the interference by $s^{-\alpha}$.}
\label{fig:tri_tdma}
\end{figure}

\fi

\end{document}